\documentclass[a4paper,11pt]{article}
\pdfoutput=1 

\usepackage{jheppub} 

\usepackage[T1]{fontenc} 
 \usepackage{etoolbox}
    \makeatletter
    \patchcmd{\maketitle}{\@fpheader}{}{}{}
    \makeatother

\usepackage{graphicx}
\usepackage[space]{grffile}
\usepackage{latexsym}
\usepackage{textcomp}
\usepackage{longtable}
\usepackage{multirow,booktabs}
\usepackage{amsfonts,amsmath,amssymb}
\usepackage{natbib}
\usepackage{url}
\usepackage{hyperref}
\usepackage{mathrsfs}
\hypersetup{colorlinks=false,pdfborder={0 0 0}}

\def\mz{M_Z^2}
\def\MZ{M_Z}
\def\al{\alpha}

\def\az{\alpha(\MZ)}

\def\dah{\Delta\alpha^{(5)}_{\rm had}}
\def\dahs{\Delta\alpha^{(5)}_{\rm had}(s)}

\def\dah0{\Delta\alpha^{(5)}_{\rm had}(-s_0)}
\def\amuh{a_\mu^{\rm had}}
\newcommand{\nn}{\nonumber}
\newcommand{\mbo}[1]{$#1$ }

\newcommand{\epm}{e^+e^-}
\newcommand{\gv}{\mbox{GeV}}

\newcommand{\mv}{\mbox{MeV}}

\newcommand{\D}{\mathrm{d}}

\newcommand{\pvint}{{\cal P} \!\!\!\!\!\!\!\! \int}
\newcommand{\semis}{\;;\;\;}
\newcommand{\bea}{\begin{eqnarray}}
\newcommand{\eea}{\end{eqnarray}}
\newcommand{\ba}{\begin{eqnarray}}
\newcommand{\ea}{\end{eqnarray}}
\newcommand{\bary}{\begin{array}}
\newcommand{\eary}{\end{array}}

\newif\iflatexml\latexmlfalse
\DeclareGraphicsExtensions{.pdf,.PDF,.png,.PNG,.jpg,.JPG,.jpeg,.JPEG}

\usepackage[utf8,cp1250]{inputenc}
\usepackage[ngerman,greek,english]{babel}

\title{Physics behind precision}

\author[a,1]{P.Azzi
\note{Editor of this document}}
\author[b]{, P. Azzurri}
\author[c]{, S. Biswas}
\author[d,1]{, F. Blekman}
\author[e]{, G. Corcella}
\author[f]{, S. De Curtis}
\author[g]{, J. Erler} 
\author[h]{, N. Foppiani }
\author[i]{, I. Helenius}
\author[j]{, S. Jadach}
\author[k]{, P. Janot}
\author[l]{, F. Jegerlehner}
\author[m,n]{, P. Langacker}
\author[o,1]{, E. Locci}
\author[p,q]{, F. Margaroli}
\author[q]{, B. Mele}
\author[r,1]{, F. Piccinini }
\author[s]{, J.Reuter}
\author[t]{, M. Steinhauser}
\author[b,1]{, R. Tenchini}
\author[u]{, M. Vos}
\author[v]{, C. Zhang}

\affiliation[a]{ INFN, Sezione di Padova, Italy}
\affiliation[b]{INFN, Sezione di Pisa, Italy}
\affiliation[c]{ KIAS, 85 Hoegi-ro, Dongdaemun-gu, Seoul 130-722, Republic of Korea} 
\affiliation[d]{IIHE, Vrije Universiteit Brussel, Pleinlaan 2, 1050 Brussel, Belgium}
\affiliation[e]{INFN, Laboratori di Frascati, Roma, Italy}
\affiliation[f]{INFN and Universit\'a di Firenze, Firenze, Italy} 
\affiliation[g]{Instituto de F\'isica, Universidad Nacional Aut\'onoma de M\'exico, Apartado Postal 20--364, M\'exico D.F. 01000, M\'exico}
\affiliation[h]{Scuola Normale Superiore, Pisa, Italy}
\affiliation[i]{Department of Astronomy and Theoretical Physics, Lund University, S\"{o}lvegatan 14A,SE-223 62 Lund, Sweden}
\affiliation[j]{Department of Particle Theory, Institute of Nuclear Physics, ul. Radzikowskiego 152, 31-342  Cracow,  Poland} 
\affiliation[k]{ CERN, EP Department, Geneva, Switzerland}
\affiliation[l]{ Institut f\" ur Physik Humboldt-Universit\" at zu Berlin, Newtonstra\ss e 15, D-12489 Berlin, Germany} 
\affiliation[m]{ Princeton University, Princeton, NJ 08544, USA} 
\affiliation[n]{Institute for Advanced Study, Princeton, NJ 08540, USA} 
\affiliation[o]{ CEA/IRFU, Saclay, France}
\affiliation[p]{ Dipart. di Fisica, Sapienza Universit\`a di Roma, Piazzale Aldo Moro 2, I-00185 Rome, Italy}  
\affiliation[q]{INFN, Sezione di Roma, Piazzale Aldo Moro 2, I-00185 Rome, Italy} 
\affiliation[r]{ INFN, Sezione di Pavia, Italy}
\affiliation[s]{ DESY, Notkestr. 85, D-22603 Hamburg, GERMANY} 
\affiliation[t]{ Karlsruhe Institute of Technology (KIT), D-76128 Karlsruhe, Allemagne}
\affiliation[u]{ IFIC, Poligono la Coma, Paterna, Valencia, Spain} 
\affiliation[v]{ Department of Physics, Brookhaven National Laboratory, Upton, NY 11973, USA}

\abstract{This document provides a writeup of contributions to the FCC-ee mini-workshop on "Physics behind precision" held at CERN, on 2-3 February 2016. }

\begin{document}
\maketitle
\flushbottom
\section{Foreword}
\label{sec:Foreword}
This document presents the highlights of the 10th FCC-ee mini-workshop
``Physics behind precision'', held at CERN on 2-3 February 2016~\cite{indico}.
The main purpose of the mini-workshop was to review the precision goals of FCC-ee
for Z peak, W pair threshold and top-quark physics
and to discuss the physics reach of precision
measurements in comparison with conceivable BSM scenarios. 
The workshop schedule included seventeen talks, which are summarized in the
following pages.

The workshop opened with an historical perspective of the role played by precision
at $e^+ e^-$ colliders in testing the Standard Model and constraining New Physics (beyond
the Standard Model) having in mind possible scenarios at future $e^+ e^-$
colliders. The subsequent talks focussed on few key items at future $e^+ e^-$ machines,
related to the precision as well as to the BSM potential, in view of the conceivable
accuracy of the measurements and theoretical calculations. On the precision side,
the following topics have been discussed: 
\begin{itemize}
\item  electroweak physics at the Z peak, with particular attention to the challenges 
   posed by the uncertainties related to the photon vacuum polarization; 
\item  recent developments in the study of the W mass and width determination through an
energy threshold scan; 
\item present accuracy in the knowledge of the $t \bar t$ threshold at FCC-ee and
the top quark mass determination through the energy threshold scan.
\end{itemize}

The potential of Z- and W-boson physics with a statistics four to five times larger than the one collected at LEP has been clearly underlined at the workshop.
The physics of the top quark has been the link between the precision and what
can be reached behind precision. A session was devoted to BSM physics reach
with top quark pairs in scenarios of composite Higgs models. Also possible
BSM scenarios involving FCNC single top processes has been addressed. 
A common theme of the discussions has been the critical comparison of the reach
among different machines, such as FCC-ee, FCC-hh, ILC, HL-LHC.

\section{Precision Measurements \& Their Sensitivity to New Physics}
{\it Paul Langacker}

Precision  electroweak studies~\cite{ALEPH:2005ab,Schael:2013ita,Langacker:1226768} have played a major role in establishing the standard
model at both the tree  (gauge principle, group, representations) and loop  (renormalization theory, predictions of the top and Higgs masses) levels. They have also significantly constrained the possibilities for new physics beyond the standard model and allowed precise measurements of the gauge couplings, suggesting supersymmetric gauge unification.

The discovery of the Higgs boson completes the verification of the standard model as a mathematically consistent
theory that successfully describes most aspects of nature down to $10^{-16}$ cm, a feat that was almost undreamt of 50 years ago. Nevertheless, the standard model is very complicated, involves severe fine-tunings, has many free parameters, and does not address such questions as the existence of three families. It also does not contain the
dark matter and energy, explain the excess of baryons over antibaryons, or incorporate  quantum  gravity.
It can be extended to include either Dirac or Majorana neutrino masses, but we do not know which.

The observation of the Higgs-like boson raises new questions. Its 125 GeV mass is rather low if there is no physics beyond the standard model up to high scales, implying a metastable vacuum. Conversely, it is rather high for the
minimal supersymmetric extension  of the standard model. Moreover, the quadratic divergences in the
Higgs mass-square suggest that the natural value should be  the Planck scale, some 17 orders of magnitude above what is 
observed. To avoid extreme fine-tuning, naturalness arguments have long been invoked to predict that the divergences would be cancelled or cut off by new physics at the TeV scale, such as supersymmetry, strong dynamics, or extra space dimensions. However, 
there is so far no unambiguous evidence for such effects from the LHC.  It may be that the new physics is just around the corner and will show up in Run 2, or perhaps at the FCC-hh. However, it is also possible that the naturalness
paradigm is not correct, perhaps replaced by environmental considerations (such as have often been invoked for the even
more unnatural value of the vacuum energy). 

It is perhaps time to examine two other paradigms as well: Are the laws of physics unique, perhaps controlled by  some elegant underlying symmetry, or are they in part determined by our accidental location in a vast multiverse of vacua, as suggested by string theory and inflation? Is new physics minimal, as usually assumed in bottom-up models, or does it
involve remnants (such as $Z'$ gauge bosons, additional
Higgs doublets or singlets, and new vector-like fermions) that are not motivated by any particular low-energy problem, but which instead accidentally remained light in the breaking of  high-scale symmetries?

There are a number of ways (not all mutually exclusive) in which nature may choose to address the shortcomings of the standard model
and the paradigms of naturalness vs tuning, uniqueness vs environment, and minimality vs remnants.
These include
\begin{itemize}
\item Strong dynamics at some relatively low scale, such as in composite Higgs models.
\item A low fundamental scale, such as large or warped dimensions or a low string scale.
\item A perturbative connection to a very high scale, such as in string theory or grand unification.
Such theories may involve (possibly nonminimal) supersymmetry or other such physics, remnants, or perhaps nothing new.
\item A vast multiverse of vacua, which could include all of the above.
\end{itemize}

The precision and Higgs programs at the FCC-ee would be extremely sensitive  to many of the possible new particles and related effects. If observed the next difficulty might be in distinguishing, e.g., between new particles from strong dynamics or string remnants. A balanced program of complementary experiments, including, for example,
the FCC-ee and FCC-hh would be essential.

\section{Status of Top Physics} 
{\it Marcel Vos}
\subsection{Introduction}

\begin{figure}[h!]
\begin{center}
\includegraphics[width=0.70\columnwidth]{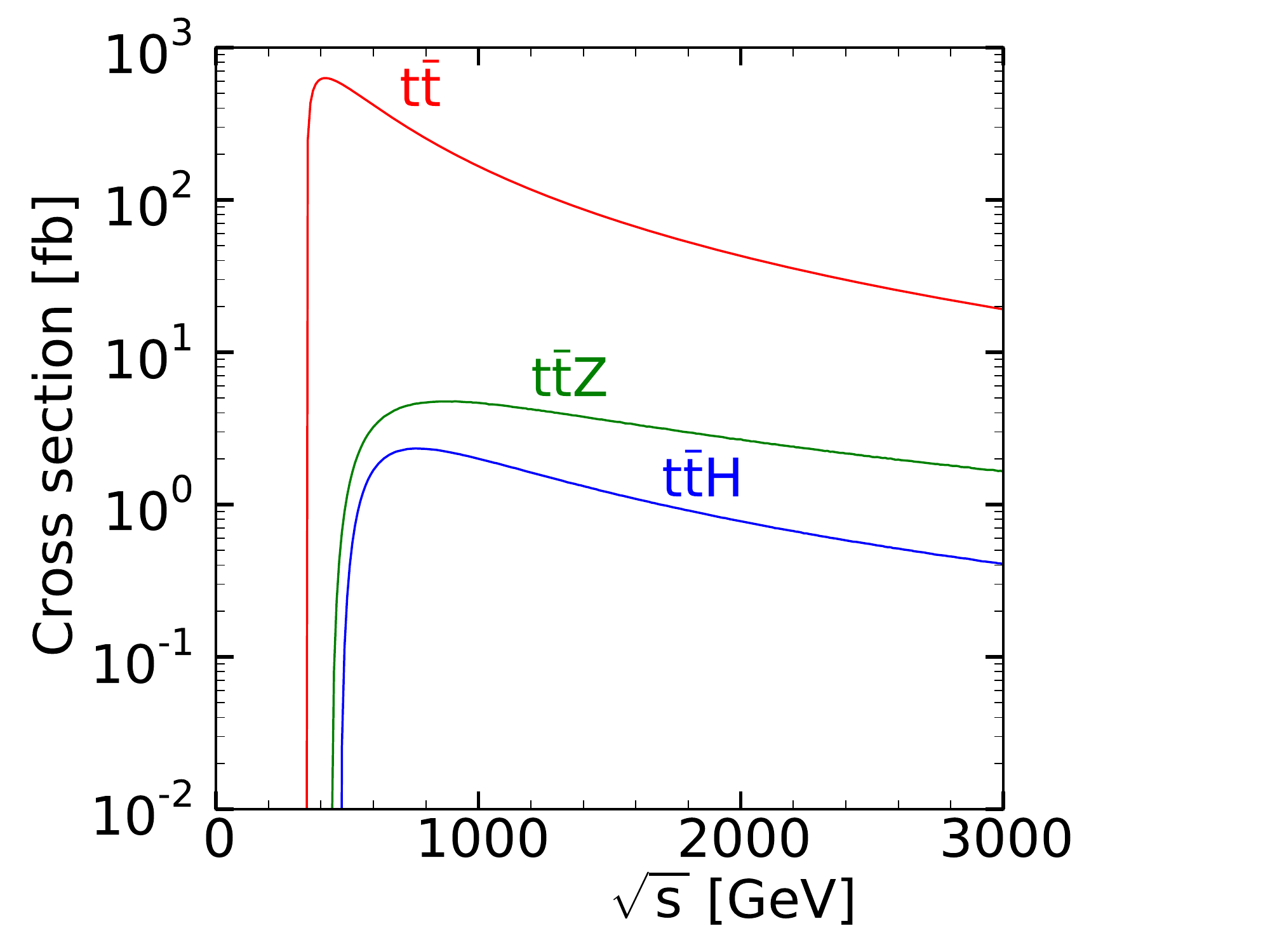}
\caption{{The production cross section versus center-of-mass energy for top quark pair production (upper, red curve) and associated production of a top quark pair and a $Z$-boson (central, green curve) or a Higgs boson (lower, blue curve).
\label{fig:processes}%
}}
\end{center}
\end{figure}

Top quarks have never been produced at lepton colliders. Lepton colliders with
sufficient energy therefore open up a completely new set of measurements. 
The evolution of the cross section versus center-of-mass energy is 
shown in Figure~\ref{fig:processes}

The first threshold, that of top quark pair production, lies at a 
center-of-mass energy equal to twice the top quark mass, i.e. 
$\sqrt{s} \sim 2 m_t \sim$ 350\,GeV. The second important threshold, 
that of top quark pair production in association with a Higgs boson, lies at 
$\sqrt{s} \sim$ 500\,GeV. Yet higher energies may be required to
produce undiscovered particles decaying to top quarks and provide exquisite 
sensitivity to new massive mediators in top quark production even
if they are outside the direct reach of the machine.

The largest among the circular 
$e^+e^-$ colliders currently under consideration (TLEP or 
FCC-ee~\cite{Gomez-Ceballos:2013zzn}) could possibly
reach the top quark pair production theshold with a non-negligible 
luminosity~\cite{Janot:2015yza}. All $e^+e^-$ colliders envisage a scan
through the pair production threshold.
Higher energies can be explored with a long linear collider. The ILC 
project~\cite{Fujii:2015jha,Baer:2013cma,Behnke:2013lya} furthermore 
envisages a main stage with large 
integrated luminosity (up to 2.6\,ab$^{-1}$~\cite{Barklow:2015tja}) 
at $\sqrt{s} = 500$\,GeV and is upgradeable to 1\,TeV. 
A relatively small investment to raise the center-of-mass energy to
550\,GeV yields a factor two improvement in the measurement of the
Yukawa coupling and should definitely be pursued~\cite{Fujii:2015jha}. 
The CLIC project~\cite{Linssen:2012hp,Aicheler:2012bya} for a relatively
compact collider with multi-TeV capacity envisages a first low-energy 
stage~\cite{CLIC:2016zwp} at 380\,GeV.

Two further projects exist, that will not be discussed. The Chinese 
CEPC~\cite{CEPC-SPPCStudyGroup:2015csa}
focuses on operation at $\sqrt{s} = $ 250\,GeV, the energy that maximizes
the cross section for the Higgs-strahlung process. Its pre-CDR does not envisage
operation at $\sqrt{s} >2 m_t$. A high-energy muon
collider~\cite{Alexahin:2013ojp} remains an interesting option, but its 
physics potential has not been explored yet.  

Several reports from strategy studies~\cite{Agashe:2013hma} or dedicated 
workshops~\cite{Vos:2016til} provide a more or less up-to-date description 
of top quark physics of future lepton colliders.

\subsection{Top quark mass}

A threshold scan through the pair production threshold is a unique opportunity 
to measure top quark properties. The line shape of the threshold depends 
strongly on the top quark width and mass~\cite{Gusken:1985nf}. The latter
is a key input to the SM and a precise measurement is crucial to test the
internal consistency of the theory~\cite{Baak:2014ora,Degrassi:2012ry}.
The rate right above threshold moreover has a strongly enhanced sensitivity 
to the strong coupling constant of QCD and the Yukawa coupling 
of the top quark.

The theory calculations required for an analysis of the threshold shape
have a surprising level of maturity: calculations have reached NNNLO
precision in QCD~\cite{Beneke:2015kwa} (or, alternatively, 
NNLL~\cite{Hoang:2013uda})
and include a host of electro-weak and off-shell corrections. The
conversion between threshold mass schemes and $\overline{MS}$ is known to
four loops~\cite{Marquard:2015qpa}.

A fit to a threshold scan, that would take well under a year to perform,
can extract the top quark mass with a statistical uncertainty of order
20\,MeV (small differences between Ref.~\cite{Seidel:2013sqa,Horiguchi:2013wra,Martinez:2002st} can be understood in terms of assumptions on the beam 
polarization and details of the fit). The largest systematic uncertainties
are expected to be due to scale uncertainties~\cite{Simon:2016htt} and 
the conversion to $\bar{MS}$ scheme~\cite{Vos:2016til}. A total uncertainty
of 50\,MeV is a realistic prospect at any lepton collider capable of
collecting 100\,fb$^{-1}$ at threshold\footnote{A target of 10\,MeV  
precision has been reported in some documents; while it is clear that
such a statistical precision can be reached at the lepton collider
projects considered here it is equally clear that the total uncertainty
will be dominated by other sources unless a theory breakthrough is made.}.

At the LHC the experimental uncertainty is expected decrease significantly
in the next decades~\cite{Juste:2013dsa,CMS-PAS-FTR-13-017}, to 
200-500~\,MeV{}. The future evolution of the ucertainty due to the
ambiguity in the interpretation of the most precise measurements
(currently estimated as ${\cal O} (1\,GeV)$~\cite{Butenschoen:2016lpz,Juste:2013dsa,ahoang08,Moch:2014tta,ahoang14,Corcella:2015kth}) 
 is much harder to predict and the perspective for the ultimate precision
at hadron colliders remain unclear~\cite{Mangano:2016jyj}.

\subsection{Top and EW gauge bosons}

The couplings of the top quark to neutral electro-weak gauge are relatively
unconstrained by experiment. A precise characterization of the $t\bar{t}Z$ 
and $t\bar{t}\gamma$ vertices is possible at lepton colliders, where 
$e^+e^- \rightarrow Z/\gamma^*$ is the dominant production process.
Several studies, either using a simple fit on cross section and 
forward-backward asymmetries measured with different beam 
polarizations~\cite{Amjad:2013tlv,Amjad:2015mma} 
or with a more sophisticated matrix-element 
fit~\cite{Janot:2015yza,Khiem:2015ofa},
show that the relevant form factors can be constrained to the \%-level.
The measurement of the CP-odd triple products proposed in 
Ref.~\cite{bib:cpvbernreuther2} yields tight constraint on CP violation
in the top quark sector~\cite{AguilarSaavedra:2001rg}.

The first preliminary studies of the sensivity versus center-of-mass energy 
show that the axial coupling to the $Z$-boson is best constrained at energies
well above the threshold, as the dependence scale with the top quark velocity.
At energies beyond a \,TeV{} the sensivity decreases for most form factors,
as a result of the decreasing production rate. It increases, however,
for the dipole moments 

Key to this precision is the predictability of rates and distributions at
a lepton collider. QCD corrections to the pair
production cross section at NNNLO are estimated at a few per mil for 
$\sqrt{s} \sim $ 500~\,GeV. Further work is needed to estimate the uncertainty 
due to uncalculated NNLO EW corrections and the uncertainty
close to threshold, where bound-state where the sensitivity to QCD 
and EW scale variations is strongly enhanced. 

The constraints that can be derived from experiments at any of the $e^+e^-$
colliders considered here are an order of magnitude tighter than those

\subsection{Top and Higgs}

The interaction of the golden couple formed by top quark and Higgs bosons 
is arguably one of the most intriguing areas of particle physics today.
The precision to which the strength of the top quark Yukawa coupling 
can be measured in $t\bar{t}H$ production 
has been studied by CLIC\cite{Abramowicz:2016zbo} and 
ILC~\cite{Price:2014oca,Fujii:2015jha}. Both projects expect 3-4\% precision
in a broad range of center-of-mass energies: $550 < \sqrt{s} < 1.5$ \,TeV.
This precision is competitive compared to the order 10\% precision of the 
HL-LHC programme~\cite{Dawson:2013bba}.

\subsection{FCNC top interactions}

Flavour Changing Neutral Current interactions with top quarks are strongly
suppressed in the Standard Model, but several extensions of the SM predict 
branching ratios $t \rightarrow qZ$, $t \rightarrow q\gamma$, $t \rightarrow qh$ of order $10^{-5}$ that might be observable at colliders.    
Lepton colliders can provide competitive constraints in the case of 
$t\rightarrow ch$, as the dominant $h\rightarrow b\bar{b}$ decay is 
more readily distinguished from background than at hadron colliders and
charm tagging achieves better performance. 

A parton-level study~\cite{Zarnecki_ichep} indicates that an $e^+e^-$ collider 
with large integrated luminosity (greater than 500~\,fb$^{-1}$) can probe the 
FCNC decay $t\rightarrow ch$ to a branching ratio of approximately $10^{-5}$,
well in excess of the HL-LHC prospects~\cite{Agashe:2013hma}.

\subsection{Summary and outlook}
\label{sec:summary}

Future high-energy $e^+e^-$ colliders offer a unique chance for precision
physics of the Higgs boson and top quark, comparing SM predictions and 
measurements at the per-mil level. The prospects for a selection of key
measurements are compared to those of the HL-LHC programme in Table~\ref{tab:summary}.

\begin{table}[h]
\label{tab:summary}
\begin{tabular}{l c c c} \\ \hline
measurement     & HL-LHC  & $e^+e^-$  & data set  \\    
                     &         &          & \\   \hline
mass (exp.)  & 200\,MeV & 20\,MeV~\cite{CMS-PAS-FTR-13-017}   & 100\,ab$^{-1}$ at $\sqrt{s} \sim 2 m_t$ \\
   $\oplus$   (theo.)        &     ?*         &        50\,MeV & ~\cite{Vos:2016til,Seidel:2013sqa,Horiguchi:2013wra,Martinez:2002st,Simon:2016htt} \\ 
            &  & & \\         
EW couplings & &  & 0.5~\,ab$^{-1}${}, 500~\,GeV~\cite{Amjad:2013tlv,Amjad:2015mma,Khiem:2015ofa} \\
 $F_{1V/1A/2V/2A}^{\gamma/Z}$   &  0.1 \cite{Rontsch:2016gqe,Rontsch:2015una,Baur:2004uw}   & 0.01**   &  0.5~\,ab$^{-1}${}, 380~\,GeV~\cite{CLIC:2016zwp} \\ 
                                 &   &  &  2.6~\,ab$^{-1}${}, 360~\,GeV~\cite{Janot:2015yza} \\
BR($t \rightarrow ch$)  & $10^{-4}$   ~\cite{Agashe:2013hma}   &  $10^{-5}$  & 1~\,ab$^{-1}${}, 380-500~\,GeV  ~\cite{Vos:2016til,Zarnecki_ichep}\\    
                       &         &          & \\   
Yukawa & 10\%  \cite{Dawson:2013bba} & 3-4\% & 1~\,ab$^{-1}${}, 0.55 - 1.5~\,TeV \cite{Fujii:2015jha,Price:2014oca,Abramowicz:2016zbo} \\ \hline 
\end{tabular}
\caption{{Brief summary of the prospects of the LHC programme, including the luminosity upgrade (3~/iab{} at 14\,TeV), and those of the different electron-positron projects for top quark physics.  The ambiguity in the interpretation of direct mass measurement is currently estimated at order 1~\,GeV~\cite{Butenschoen:2016lpz,Juste:2013dsa,ahoang08,Moch:2014tta,ahoang14,Corcella:2015kth}. 
Progress in this area is hard to predict~\cite{Mangano:2016jyj}. The constraints depend on the center-of-mass energy in a non-trivial way, with axial couplings constrained best when the top quarks have velocity $\beta \sim $ 1 ($\sqrt{s} > $ 500~\,GeV). Four-fermion operators are best constrained at very high energy. }}
\end{table}

A scan of the beam energy through the top quark pair production 
threshold ($\sqrt{s} \sim 2 m_{t}$) allows for an extraction of the $\bar{MS}$ 
mass to a precision of approximately 50\,MeV (including all theoretical
uncertainties). Precision studies of pair production above threshold
constrain the couplings to neutral electro-weak gauge bosons to the \%-level,
exceeding the current precision by two orders of magnitude and that 
envisaged after 3~\,ab$^{-1}${} at the LHC by one order. A direct measurement of
the $t\bar{t}H$ production rate at a center-of-mass energy greater than 
500~\,GeV yields a competitive determination of the top Yukawa coupling,
with a precision of 3-4\%.


\section{Precise measurements of the top mass: theory vs.~experiment}
\label{Corcella}
{\it Gennaro Corcella}

The top quark mass ($m_t$) is a fundamental parameter of the Standard Model, 
constraining the Higgs boson mass even before its
discovery; it plays a role in the result that the electroweak vacuum
lies at the border between metastability and instability regions
\cite{Degrassi:2012ry}.
This statement depends on the precision achieved on the $m_t$ determination,
about 700 MeV in the world average \cite{ATLAS:2014wva}
and 500 MeV in the latest CMS analysis \cite{Khachatryan:2015hba}, 
as well as on the assumption that the measured value
corresponds to the pole mass. 

Standard hadron-collider techniques,  
such as the template or matrix-element methods,
use Monte Carlo generators, namely 
parton showers and hadronization 
models, which are not exact QCD calculations. However, the very fact that
such measurements 
reconstruct top-decay ($t\to bW$) final states, with on-shell
top quarks, makes the identification of the measured $m_t$ with
the pole mass quite reasonable, up to theoretical uncertainties, such as
missing higher-order and width contributions, or colour-reconnection
effects. A recent analysis, based on the 4-loop calculation of
the relation between pole and $\overline{\rm MS}$ masses 
\cite{Marquard:2015qpa} and its comparison with
the renormalon computation \cite{Beneke:1994sw}, showed that 
the renormalon ambiguity on the
pole mass is even below 100 MeV \cite{Nason:2016tiy}, 
thus making it a reliable quantity.
Furthermore, studies 
carried out in the SCET framework found that the discrepancy
between the pole and the so-called `Monte Carlo mass' is about 200 MeV
\cite{Hoang:2008xm}.
Work is presently undertaken to assess the non-perturbative uncertainty
on $m_t$, due to colour reconnection, by comparing standard $t\bar t$ events
with final states obtained from fictitious $T$-meson decays \cite{Corcella:2014rya,Corcella:2015kth}. 

Within the so-called alternative methods, the total $t\bar t$ 
\cite{CMS:2015uoa,Aad:2014kva} and 
$t\bar t$+jet cross section \cite{Aad:2015waa}, calculated respectively in 
the NNLO \cite{Czakon:2013goa}
and NLO \cite{Alioli:2013mxa} 
approximations, have been used to extract the pole mass. 
The current uncertainty, using the LHC Run I data, is about 2 GeV
and should go down to $\sim$~1 GeV, thanks to the higher Run II statistics.
Other strategies, based on endpoints \cite{Chatrchyan:2013boa}, 
$b$-jet energy peaks \cite{Agashe:2016bok,CMS:2015jwa}, $b$-jet+$\ell$ 
invariant mass \cite{CMS:2014cza} and
$J/\psi+\ell$ final states \cite{Chierici:2006dh}, 
are very interesting, but
will not yield measurements more precise than those achieved with
standard methods. 
 
At future $e^+e^-$ colliders, the threshold scan of the $t\bar t$ production
cross section $\sigma_{t\bar t}$ allows a very precise determination of $m_t$:
in this regime,
the 1S \cite{Hoang:1999zc} and the potential-subtracted (PS) masses \cite{Beneke:1998rk}
are adequate definitions.
The ratio $R=\sigma_{t\bar t}/\sigma_{\mu^+\mu^-}$ is typically expressed as 
a series of powers of $\alpha_S$ and $v$, the velocity of the top quarks:
\begin{equation}
R=
v\sum\limits_k\left(\frac{\alpha_S}{v}\right)^k\sum\limits_i
\left(\alpha_S\ln v\right)^i
\left\{1({\rm LL}); \alpha_S,v ({\rm NLL});
\alpha_S^2,\alpha_Sv, v^2({\rm NNLL})\right\}.\end{equation}
The state of the art of the calculation of $R$ is NNNLO \cite{Beneke:2015kwa}, 
where 
the N$^k$LO approximation includes contributions  $\alpha_S^mv^n$,
with $m+n=1,\dots k+1$, and NNLL \cite{Hoang:2013uda}, 
where terms $\alpha_S^k\ln^kv$ are LL, 
$\alpha_S^{k+1}\ln^kv$ and $v\alpha_S^k\ln^kv$ are NLL and so on.
The NNNLO computation uses the PS-subtracted mass and yields 
$\Delta m_t^{\rm PS}\leq 50$~MeV, when
extracting $m_t^{\rm PS}$  from the peak of $R$. The NNLL resummation
employs the 1S mass and leads to $\Delta m_t^{\rm 1S}\simeq 40$~MeV.

By carrying out a threshold scan of $R$ at
${\cal L}=100$~fb$^{-1}$, in the CLIC, ILC and FCC-$ee$ scenarios and using the NNLO+NNLL TOPPIK
code $m_t^{1S}$ can be determined through a 1D or a 2D fit, according to wether $\alpha_S$ is fixed or fitted\cite{Seidel:2013sqa}.
The resulting statistical errors are $\Delta m_t^{\rm stat}\simeq $~21-
34 MeV (CLIC), 18-27 MeV (ILC) and 16-22 MeV (FCC-$ee$).
The theoretical uncertainty, gauged as a 1\% error on the $R$ normalization is
$\Delta m_t^{\rm theo}\simeq$~5-18 MeV (CLIC and ILC), 8-14 MeV (FCC-$ee$);
if the theory error grows to 3\%, we have 
$\Delta m_t^{\rm theo}\simeq$~8-56 MeV (CLIC), 9-55 MeV (ILC)
and 9-41 MeV (FCC-$ee$).

\section{Top quark threshold production and the top quark mass}
{\it  Matthias Steinhauser}

In this contribution we discuss the third order QCD corrections to top quark 
threshold production and elaborate on the precision of the conversion formula 
relating the extracted threshold mass to the  top quark mass in the $\overline{\rm MS}$ scheme.

An important task of a future electron-positron collider is a precise 
measurement of the total cross section $\sigma(e^+e^-\to t\bar{t})$
in the threshold region. From comparison with theoretical predictions it is
possible to extract precise values for the top quark mass. Furthermore,
also the top quark width, the strong coupling and the top quark Yukawa coupling
can be determined to high precision.

The calculation of the threshold cross section requires that the top-anti-top
system is considered in the non-relativistic limit.  Due to the fact that both
mass and width of the top quark are quite large, perturbation theory should be
applicable. However, truncating the perturbative series at
next-to-next-to-leading order (NNLO)~\cite{Hoang:2000yr} leads to
unsatisfactory results; there is no sign of convergence in the peak
region. This has triggered a number of calculations of N$^3$LO
building blocks (see Ref.~\cite{Beneke:2013jia} for a
detailed discussion).  Among the most important ones are the three-loop
corrections to the static
potential~\cite{Smirnov:2008,Smirnov:2009,Anzai:2009} and three-loop
corrections to the vector current matching coefficient between QCD and
NRQCD~\cite{Marquard:2006,Marquard:2009,Marquard:2014}. Furthermore,
ultrasoft corrections, all Coulombic contributions up to the third order, and
all single- and double non-Coulomb potential
insertions~\cite{Beneke:2008,Beneke:2005,BenKiy_II} have been computed to
the non-relativistic two-point Green function of the top-anti-top system.

\begin{figure}[h!]
\begin{center}
\includegraphics[width=0.70\columnwidth,angle=90]{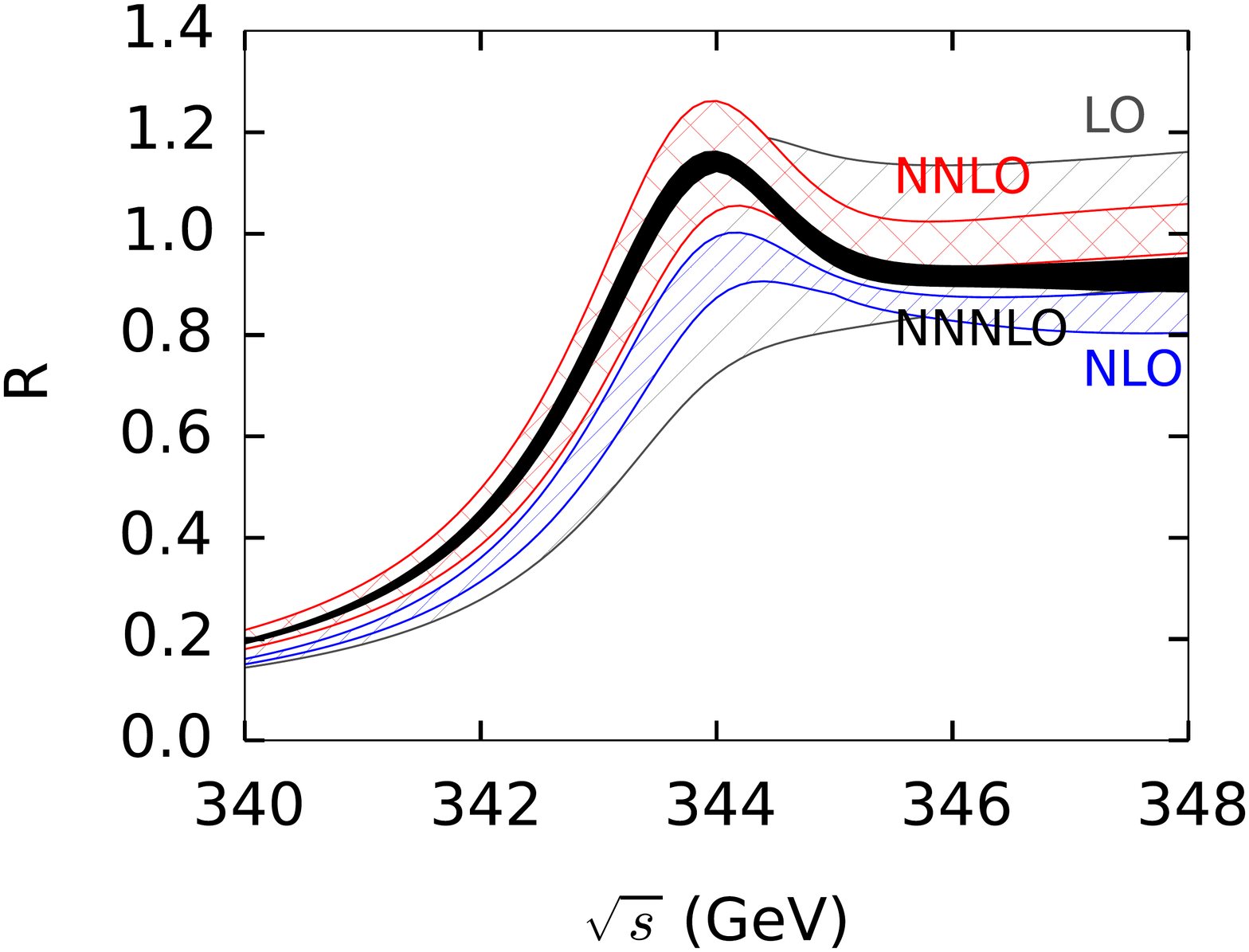}
\caption{{Scale dependence of the cross section near threshold. The LO, NLO,
    NNLO and N$^3$LO result is shown in gray, blue, red and black,
    respectively. The renormalization scale $\mu$ is varied between $50$ and
    $350$~GeV.
    \label{fig::mu_a}%
}}
\end{center}
\end{figure}

\begin{figure}[h!]
\begin{center}
\includegraphics[width=0.70\columnwidth,angle=90]{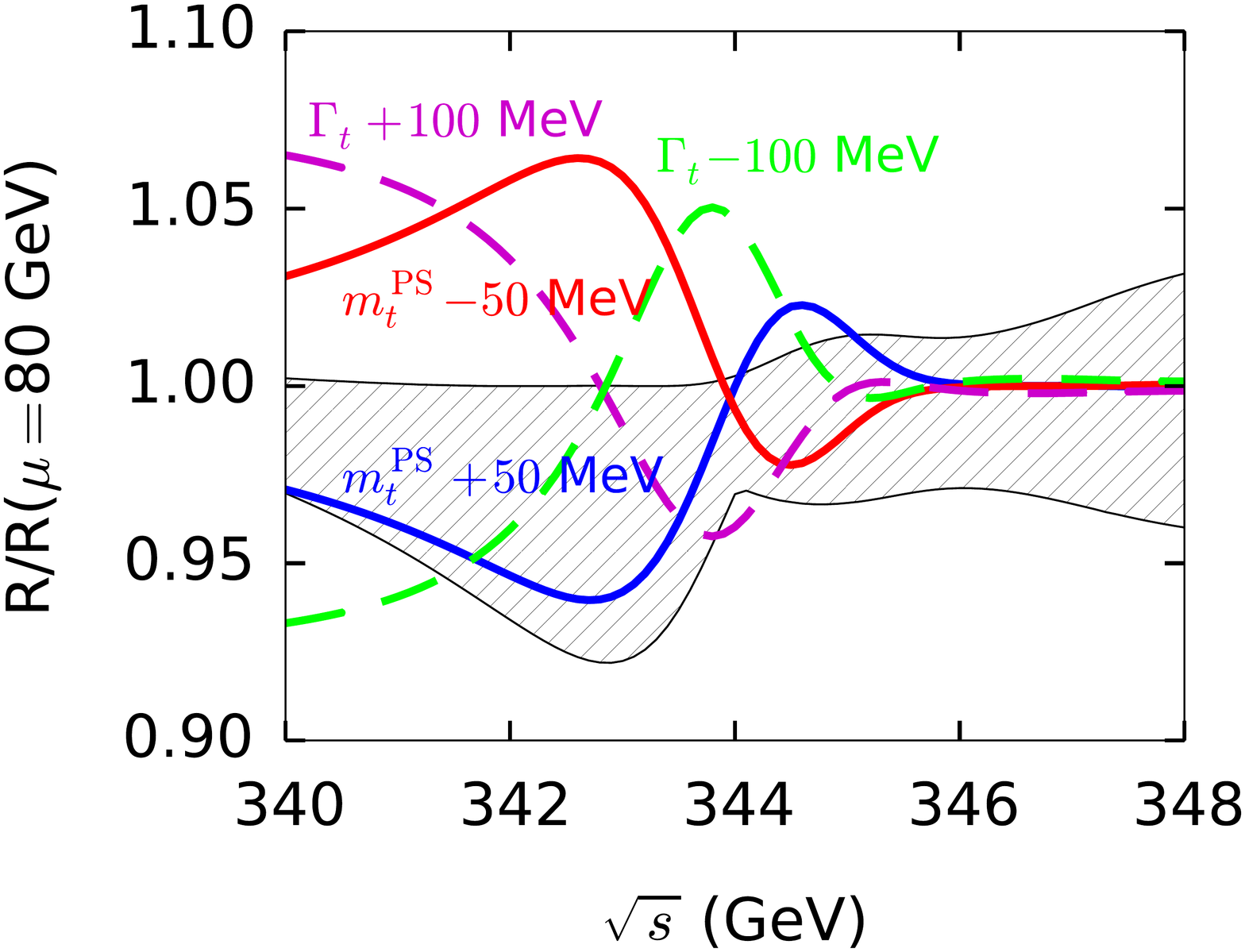}
\caption{{N$^3$LO cross section normalized to $R(s)$ for
    $\mu=80$~GeV. The solid and dashed curves correspond to a variation
    of PS top quark mass and top quark width, respectively.
    \label{fig::mu_b}%
}}
\end{center}
\end{figure}

The total cross section up to N$^3$LO (normalized to
$\sigma(e^+e^-\to\mu^+\mu^-)$) is shown in Fig.~\ref{fig::mu_a}
where the bands are obtained from variation of the renormalization scale (see
caption for details)~\cite{Beneke:2015kwa}. One observes a big jump from NLO to
NNLO. However, in and below the peak region the N$^3$LO band lies 
on top of the NNLO band and has a significantly reduced uncertainty.
The normalized N$^3$LO cross section is shown in Fig.~\ref{fig::mu_b} as
hatched band. One observes an uncertainty of about $\pm3\%$.
Fig.~\ref{fig::mu_b} also shows the effect of the variation of 
the top quark mass and width and demonstrates that the current theory
uncertainty allows for a sensitivity of $\delta m_t^{\rm PS}\approx 50$~MeV
and $\delta\Gamma_t\approx 100$~MeV.

There are various beyond-QCD effects which are available on top of the N$^3$LO
QCD corrections like electroweak corrections, non-resonant $W^+W^-b\bar{b}$
production, QED effects and $P$-wave production.  Their numerical effect is
taken into account in Ref.~\cite{Beneke:2015lwa}. However, the overall
features of the pure-QCD result (in particular the obtain precision) is not
changed.

As can be seen from Fig.~\ref{fig::mu_b}, the determination of the top quark
mass defined in a properly chosen threshold scheme with an uncertainty of
about 50~MeV is possible. In a next step the mass value has to be transformed
to the $\overline{\rm MS}$ scheme which has to done with highest possible
precision. It has been shown in Ref.~\cite{Marquard:2015qpa,Marquard:2016vmy}
that only after using the N$^3$LO formula, which is based on the four-loop
relation between the $\overline{\rm MS}$ and on-shell quark mass, a precision
of the order 10~MeV can be obtained.

\section{ Study of the sensitivity in measuring the Top electroweak couplings at the FCC-ee}
\label{sec:Foppiani}
{\it Nicol\`o Foppiani} 

The top electroweak couplings are an extremely interesting topic since they could be significantly affected by BSM physics.\\ 
However the present LHC precision of the couplings is not very constraining ($O(10\% \div 100 \%$). The scheduled FCC-ee  $4^{th}$ phase to study the Top quark is the perfect occasion to measure the top electroweak couplings with high precision.\\
The FCC-ee data can reduce the statistical uncertainties by orders of magnitude: this claim is supported, first of all, by another work (ref.~\cite{Janot:2015yza}), performed in a purely analytic way. In that paper it has been shown that it is possible to measure the top EW couplings with a satisfactory precision, with no need of incoming beam polarization. In addition it has been demonstrated that $\sqrt{s}=365$ GeV is the center of mass energy at which the statistical uncertainties of such a measurement are minimal.\\

For the results presented here, the expected precision on the measurement of the top EW coupling has been performed using fully simulated $e^+ e^-$ collisions at $\sqrt{s}=365$ GeV. The simulation, based on the CERN-CLIC group software, was obtained using the WHIZARD generator and the Marlin particle flow reconstruction algorithm \cite{Kilian:2007gr,Wendt:2007iw}. The integrated luminosity considered amounts to 2.5 $ab^{-1}$, corresponding to about 3 years of running at FCC-ee. The simulation is based on the performance of the CLIC-ILD detector, since it is currently expected that the characteristics of the FCC-ee detector will be similar, although a different technology might be used. \\

The ttV vertex (V= $\gamma$, Z) can be parametrized as in ref.~\cite{Grzadkowski:2000nx}:
\[ \displaystyle \Gamma^{\mu}_{t\bar{t}V}=\frac{g}{2}\left[ \gamma^{\mu} [(A_V + \delta A_V) - \gamma^5 (B_V + \delta B_V)]   + \frac{(p_t - p_{\bar{t}})^{\mu} }{2m_t} (\delta C_V - \delta D_V \gamma^5) \right] \]
where $A_V$ and $B_V$ are the SM couplings to the vector and axial current, whereas the $\delta X_V$ ($X=A,B,C,D$) are some possible corrections to the SM couplings. These anomalous couplings influence the top polarization which is transferred to the decay product distribution (similar to $Z \to \tau^+ \tau^-$).\\ 
So it is possible to measure the EW couplings by studying the angular-energy distribution of the lepton in the semi-leptonic decay:
\[ e^+ e^- \to t \bar t \to \bar b q q' b \ell \nu \]

The results focus on two possible corrections to the Z coupling $\delta A_Z$ and $\delta B_Z$, assuming that all the others corrections are zero. These terms, which are related to the couplings to the right and left handed part of the top quark, are nonzero in many new physics models, in particular in Composite Higgs models. 

\begin{figure}[h!]
\begin{center}
\includegraphics[width=0.70\columnwidth]{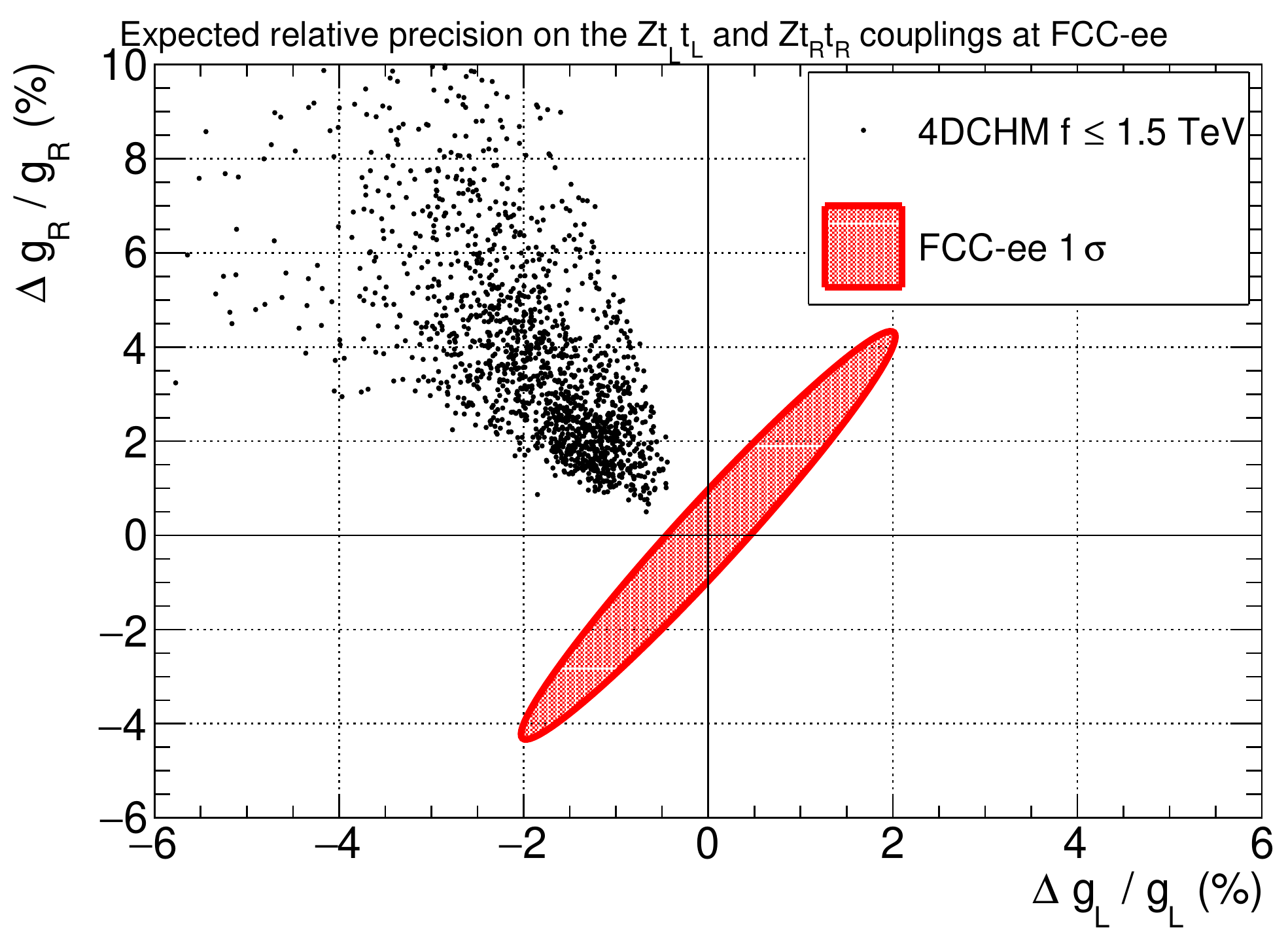}
\caption{{The expected precision on the couplings to the right and left handed part of the top quark is sufficient to exclude the 4D Composite Higgs Models with new physics energy scale lower than 1.5 TeV. The models\protect\cite{Barducci:2015aoa} that have been considered take into account also the present limits on the masses of the new resonances coming from the direct searches and from the electroweak precision test.
\label{fig:gr_gl_plot}%
}}
\end{center}
\end{figure}

The analysis starts from the samples generated with zero anomalous couplings with $\delta A_Z = \delta B_Z = 0$. The tagging efficiency of  $t \bar t$ events in the single lepton channel has been studied, including the rejection of the background, using the efficiency and purity of the selected sample as the discriminator.
The resulting distribution has been fitted with the previous distribution (after a proper smearing to take in account the detector resolution and the selection algorithm efficiency) to obtain $\delta A_Z$ and $\delta B_Z$, which are consistent with zero as expected. The uncertainties on $\delta A_Z$ and $\delta B_Z$ represent the precision that it is possible to reach with the FCC-ee and are the one standard deviation uncertainty is shown as the ellipse in figure \ref{fig:gr_gl_plot}.\\
The results confirm the analytical results in previous published work, showing that the FCC-ee could reduce significantly the statistical uncertainties on these couplings, giving a sufficient sensitivity to test some new physics models\cite{Barducci:2015aoa}, as it is shown in figure  \ref{fig:gr_gl_plot}. In fact, the relative precision obtained on the couplings to the left and the right handed top quark are of the order of few percent, enough to exclude the 4D Composite Higgs Models in this example with new physics energy scales lower than 1.5 TeV.

\section{Photon-photon interactions with {\rmfamily \textsc{Pythia}}~8: Current status and future plans}
\label{sec:Helenius}
{\it Ilkka Helenius}

Photon-photon interactions provide access to many interesting processes and serve as a further test of QCD factorization. In addition, these interactions will produce background for future $\mathrm{e^+e^-}$ colliders such as FCC-ee. To quantify the physics potential of these future experiments, accurate simulations of these background processes are required. Here we present the current status of our implementation of $\gamma\gamma$ processes into \textsc{Pythia 8} \cite{Sjostrand:2014zea} general-purpose Monte-Carlo (MC) event generator and discuss briefly about the next developments.

As with hadrons, the partonic structure of resolved photons can be described with PDFs. The scale evolution of the photon PDFs is given by the DGLAP equation
\begin{equation}
\frac{\mathrm{\partial} f^{\gamma}_i(x,Q^2)}{\mathrm{\partial}\mathrm{log}(Q^2)} = \frac{\alpha_{\rm EM}}{2\pi}e_i^2 P_{i\gamma}(x) + \frac{\alpha_s(Q^2)}{2\pi} \sum_j \int_x^1\frac{\mathrm{d}z}{z}\, P_{ij}(z)\, f_j(x/z,Q^2).
\label{eq:gammaDGLAP}
\end{equation}
The solution for this equation can be divided into a point-like part and a hadron-like part. The first describes the partons arising from $\gamma\rightarrow q\bar{q}$ splittings and can be calculated directly using pQCD. The latter requires some non-perturbative input that is typically obtained from vector meson dominance (VMD) model. In these studies we have used PDFs from CJKL analysis \cite{Cornet:2002iy}. 

To extend the \textsc{Pythia 8} parton shower algorithm \cite{Sjostrand:2004ef} to accommodate also photon beams, a term corresponding to the $\gamma\rightarrow q\bar{q}$ splitting of the original beam photon is added as a part of the initial state radiation (ISR). Since the photons do not have a fixed valence content, some further assumptions for the beam remnant handling \cite{Sjostrand:2004pf} is also required. Here we first decide whether the parton taken from the beam was a valence parton using the PDFs. If it was, the beam remnant is simply the corresponding (anti)quark and if the parton was a sea quark or a gluon, the valence content is sampled according to the PDFs. It may also happen that the hard interaction does not leave enough energy to construct the beam remnants with massive partons. Also the ISR can end up at a state where the remnants cannot be constructed due to the lack of energy. These cases (few but not negligible) are rejected in order to generate only physical events. 

The described modifications allow one to generate hard-process events in resolved interactions of two real photons with parton showers and hadronization. These developments were recently included into the public version of \textsc{Pythia 8}. Currently we are working to include also photon emissions from electrons, which are required to obtain the correct rate of $\gamma\gamma$ events in $\mathrm{e^+e^-}$ collisions at a given energy. In general the form of the photon flux depends on the machine. For a circular collider (such as FCC-ee) the dominant contribution comes from bremsstrahlung photons, which can be modeled using equivalent photon approximation (EPA). The PDFs for partons inside photons, which in turn arise from electrons, can be defined as a convolution between the photon flux and the photon PDFs
\begin{equation}
f_{i}^{\rm e}(x,Q^2) = \int_x^1 \frac{\mathrm{d}x_{\gamma}}{x_{\gamma}} \int_{\mu^2_{\mathrm{min}}}^{\mu^2_{\mathrm{max}}} \frac{\mathrm{d}\mu^2}{\mu^2} f_{\gamma}^{\rm e}(x_{\gamma},\mu^2) f^{\gamma}_i (x/x_{\gamma},Q^2).
\end{equation}
After the hard process is selected according to the $f_{i}^{\rm e}$ and $x_{\gamma}$ values are sampled, the $\gamma\gamma$ collision is set up and the usual partonic evolution is performed for the subsystem. Also addition of $\mathrm{e^-}/\gamma$+$\mathrm{p}$ collisions is in the works. These developments will soon be included to the \textsc{Pythia 8}.

\vspace{\baselineskip}
\noindent Work have been supported by the MCnetITN FP7 Marie Curie Initial Training Network, contract PITN-GA-2012-315877 and has received funding from the European Research Council (ERC) under the European Union's Horizon 2020 research and innovation programme (grant agreement No 668679).

\section{The WHIZARD generator for FCC[-ee] Physics}
\label{sec:Reuter}
{\it J\"urgen Reuter}
  {\em on behalf of the}  WHIZARD collaboration



\begin{figure}[h!]
\begin{center}
\includegraphics[width=0.7\columnwidth]{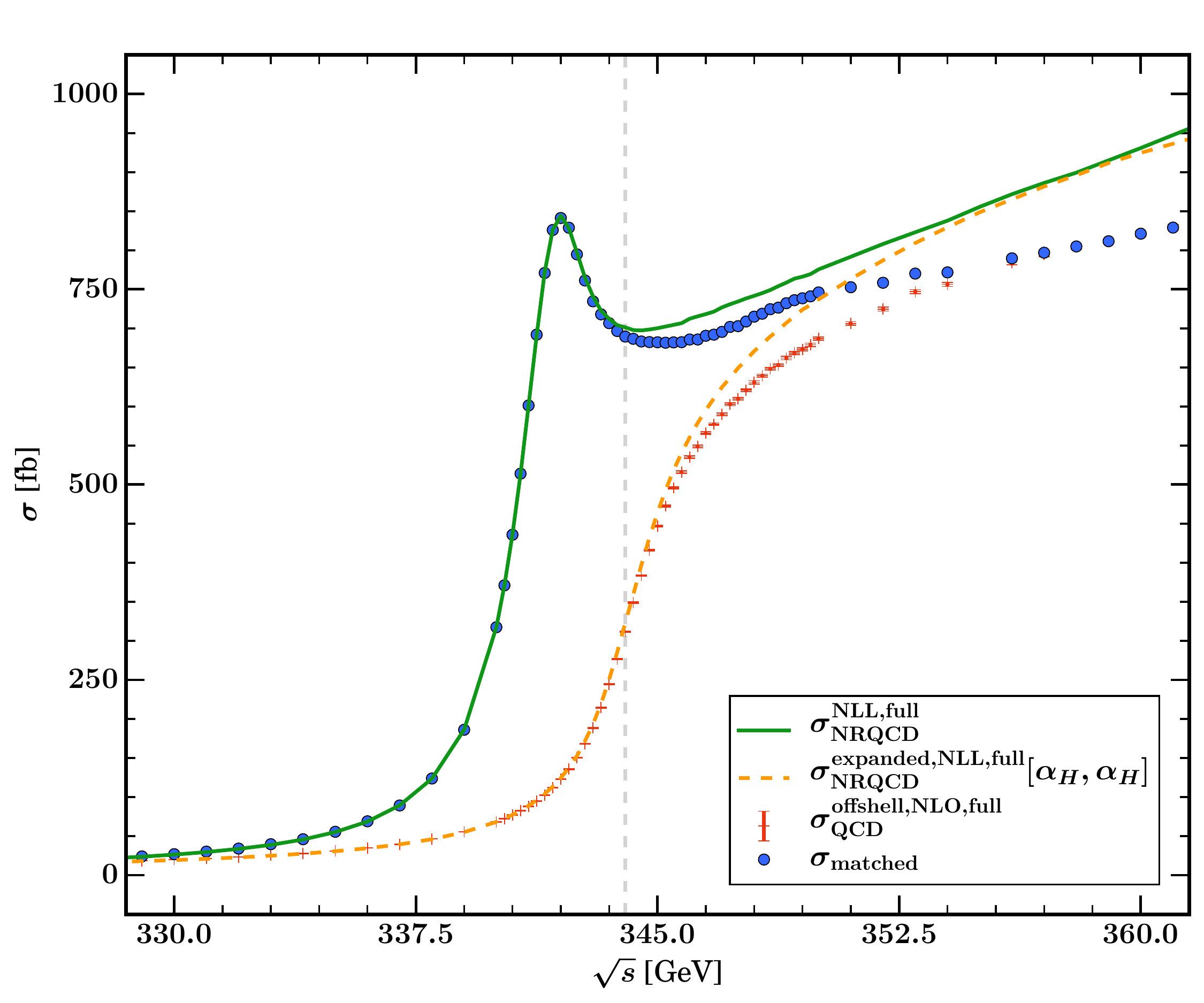}
\caption{Top threshold matching inside WHIZARD \label{fig:whizardtop}%
}
\end{center}
\end{figure}

\texttt{WHIZARD}~\cite{Kilian:2007gr} is a multi-purpose event
generator for $pp$, $ep$ and $ee$ collisions. It contains the
tree-level matrix-element generator
\texttt{O'Mega}~\cite{Moretti:2001zz} that can generate amplitudes for
many implemented models with (almost) arbitrarily high
multiplicity. New physics models can be generated via interfaces to
external programs (like \texttt{Sarah} and \texttt{FeynRules}), while
support of the UFO-file format will come in summer 2016.
processes. \texttt{WHIZARD} contains a sophisticated automated
phase-space parameterization for complicated multi-particle processes,
whose integration is performed by its adaptive multi-channel Monte
Carlo integrator, \texttt{VAMP}~\cite{Ohl:1998jn}. For lepton
colliders, important effects like beamstrahlung, beam spectra, initial
state photon radiation, polarization, crossing angles etc. are
supported. Beyond leading order, \texttt{WHIZARD} can generate events
at next-to-leading order (NLO) in the strong coupling constant for
Standard Model (SM) processes for both lepton and hadron colliders. It
automatically generates the real corrections and the subtraction terms
to render the separate contributions finite, while the virtual
amplitudes come from standard external one-loop provider programs. As
an example for a special implementation that is particularly important
for the high-energy program of the FCC-ee, Fig.~\ref{fig:whizardtop}
shows the next-to-leading logarithm top-threshold resummation matched
to the NLO continuum process. \texttt{WHIZARD} has its own module for
QCD parton showers~\cite{Kilian:2011ka} ($k_T$-ordered and analytic)
and allows for automated MLM matching to provide matched inclusive jet
samples. Only hadronization is left to external
programs. At NLO, it allows for an automated POWHEG matching to the
parton shower. \texttt{WHIZARD} can handle scattering processes also
as cascades with production and subsequent decay chains, where it
allows to keep full spin correlations, only classical ones, or to
switch them off. For polarized decays, helicities of intermediate
particles can be selected.

Besides its wide physics potential, \texttt{WHIZARD} is very
user-friendly: it allows steering of the processes, the collider
specifications, arbitrary cuts together with the analysis setup in one
single input file, using its own very easy scripting language
SINDARIN.


\baselineskip15pt

\section{$\alpha_{\rm QED}(M_Z)$ and future prospects with low energy $e^+e^-$ collider data}
\label{sec:Jegerlehner}
{\it Fred Jegerlehner}

The non-perturbative hadronic vacuum polarization contribution\\
$\dahs=-\left(\Pi'_\gamma(s)-\Pi'_\gamma(0)\right)_{\rm had}$ can be evaluated
in terms of \\
$R_\gamma(s) \equiv \sigma^{(0)}(e^+e^- \rightarrow \gamma^*
\rightarrow {\rm hadrons})/(\frac{4\pi \alpha^2}{3s})$
data and the dispersion integral
\bea 
\dahs = - \frac{\alpha\, s}{3\pi}\;\bigg(\;\;\;
 \pvint\limits_{4m_\pi^2}^{E^2_{\rm cut}} ds'
\frac{{R^{\mathrm{data}}_\gamma(s')}}{s'(s'-s)}  +\;\;\;\;\;
\pvint\limits_{E^2_{\rm cut}}^\infty ds'
\frac{{R^{\mathrm{pQCD}}_\gamma(s')}}{s'(s'-s)}\,\,
\bigg)\,.
\eea 
An up-to-date evaluation yields
$\Delta \al _{\rm hadrons}^{(5)}(\mz)=0.027648 \pm 0.000176$
and $\alpha^{-1}(\mz) = 128.942 \pm 0.021$,
where $\alpha(s)=\frac{\alpha}{1-\Delta \alpha (s)}\semis \Delta
\alpha(s)=\Delta \alpha_{\rm lep}(s)+\dahs +\Delta \alpha_{\rm
top}(s)$. Here I advocate to apply the space-like split trick which
allows us to reduce uncertainties, by exploiting perturbative QCD
in an optimal way. The monitor to control the
applicability of pQCD is the Adler function
\bea
D(Q^2=-s)=\frac{3\pi}{\alpha} s\frac{d}{\D s}\Delta \alpha_{\mathrm{had}}(s)=-(12\pi^2)\,s\,\frac{\D
\Pi'_\gamma (s)}{\D s}=Q^2\,\int_{4m_\pi^2}^{\infty}\,\frac{R(s)}{(s+Q^2)^2}
\eea
which also is determined by $R_\gamma(s)$ and can be evaluated in terms of
experimental $\epm$--data ~\cite{Eidelman:1998vc}. 

Perturbative QCD is observed to work well to predict $D(Q^2)$ down to
$M_0= 2.0\, \gv$. This may be used to separate non-perturbative and
perturbative parts by the following split of contributions:
\bea
\Delta \alpha(\mz)&=&\Delta \alpha^{\mathrm{data}}(-M^2_0)\nonumber \\
&& +\left[\Delta \alpha(-\mz)-\Delta
\alpha(-M^2_0)\right]^{\mathrm{pQCD}} \nonumber \\ &&
+\left[\Delta \alpha(\mz)-\Delta \alpha(-\mz)\right]^{\mathrm{data/pQCD}}\,,
\eea
where the space-like reference scale \mbo{M_0} is chosen such that pQCD is well
under control for \mbo{-s<-M^2_0}. $\alpha^{\mathrm{data}}(-M^2_0)$
can be evaluated by the standard approach, but evaluated in the
space-like region at fairly low scale. The second term can be obtained
by calculating
\ba
\Delta=\left[\Delta \alpha_{\mathrm{had}}(-M_Z^2)-\Delta \alpha_{\mathrm{had}}(-M_0^2)\right]^{\mathrm{pQCD}}= \frac{\alpha}{3\pi} \int_{M_0^2}^{M_Z^2}
\D Q^{2} \frac{D^{\rm pQCD}(Q^{2})}{Q^{2}}
\ea
based on the accurate pQCD prediction of the Adler function. This part
may also be computed directly as $\Delta=\left(\Pi'_\gamma(-M_0^2)-\Pi'_\gamma(-M_Z^2)\right)_{\rm pQCD}\,.$ The
third term $\Delta^{\rm reminder}=\Delta \alpha(\mz)-\Delta
\alpha(-\mz)$ is small and may be evaluated based on data in the
standard way or via pQCD, as the non-perturbative effects are expected
to cancel largely by global quark-hadron duality in the difference
between the time-like and space-like version at a high energy scale
$M_Z$. For $M_0=2.0\, \gv$ one obtains~\cite{Jegerlehner:2006ju}
\ba \bary{lcc}
\Delta\alpha^{(5)}_{\rm had}(-s_0)^{\mathrm{data}} &=& 0.006392 \pm
0.000064\\
\Delta\alpha^{(5)}_{\rm had}(-M_Z^2) &=&  0.027466 \pm 0.000118\nn \\
\Delta\alpha^{(5)}_{\rm had}(M_Z^2) &=&  0.027504 \pm 0.000119
\eary
\ea
$\Delta \al _{\rm hadrons}^{(5)}(\mz)=0.027504 \pm 0.000119$
and $\alpha^{-1}(\mz)= 128.961 \pm 0.011$,
including a shift $+0.000008$ from the 5-loop contribution
and with an error $\pm 0.000100$ added in quadrature form the perturbative part,
based on a complete 3--loop massive QCD calculation by K\"uhn et al. 2007.

The required improvement on the uncertainty of $\az$ by about a factor
5 seems to be feasible by the following means:
\begin{itemize}
\item reducing the errors in the range 1.0 to 2.5 GeV is mandatory for both
      the muon $g-2$ as well as for $\dah0$. It requires improved
      cross sections for $\sigma(\epm \to \mathrm{hadrons})$ at the
      1\% level up to energies about 2.5 GeV. 
\item the same goal can be achieved with lattice QCD calculations, which
      made big progress recently. The $\dah0$ is directly
      accessible to Lattice QCD and are expected to achieve 1\% level
      results within the next years. However, one needs an 0.4\% calculation.
\item perturbative QCD calculations of the Adler-function have to be
      extended from 3- to 4-loops full massive QCD for 5 flavors (physical
      $m_c$ and $m_b$ and common light $u, d$ and $s$ quark masses:
      $m_u\simeq m_u\simeq m_s\simeq 100~\mv$)
\item improved values for $m_c$ and $m_b$ on which the Adler function
      is rather sensitive.
\end{itemize}
Ongoing are cross section measurements with the
CMD-3 and SND detectors at the VEPP-2000 $\epm$ collider (energy scan)
and with the BESIII detector at the BEPCII collider (ISR method)

Very promising alternative possibility: determine
$\Delta\alpha^{(5)}_{\rm had}(-s_0)$ for $\sqrt{s_0}\approx 2~\gv$ 
in Bhabha--scattering as advocated for $\amuh$ in~\cite{Calame:2015fva}.

\section{Direct measurement of $\alpha_{\rm QED}(m_{\rm Z}^2)$ at the FCC-ee}
\label{sec:Janot}
{\it Patrick Janot}

When the measurements from the FCC-ee become available, an improved determination of the standard-model "input" parameters will be needed to fully exploit the new precision data towards either constraining or fitting the parameters of beyond-the-standard-model theories. Among these input parameters is the electromagnetic coupling constant estimated at the Z mass scale, $\alpha_{\rm QED}(m^2_{\rm Z})$. The measurement of the muon forward-backward asymmetry at the FCC-ee, just below and just above the Z pole, can be used to make a direct determination of $\alpha_{\rm QED}(m^2_{\rm Z})$ with an accuracy deemed adequate for an optimal use of the FCC-ee precision data.

\begin{figure}[h!]
\begin{center}
\includegraphics[width=0.98\columnwidth]{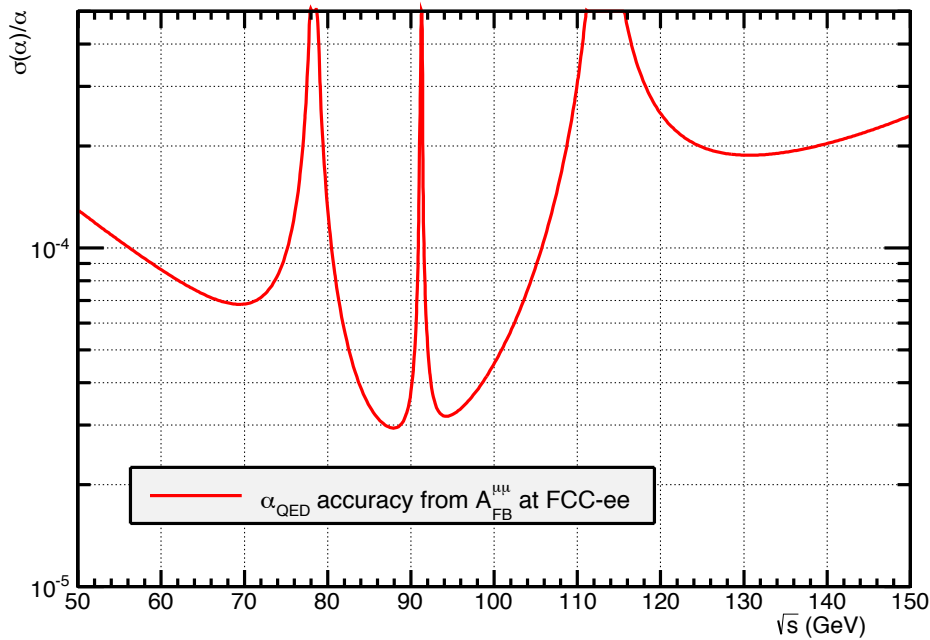}
\caption{\label{fig:sensitivity}
Relative statistical uncertainty for the $\alpha_{\rm QED}(s)$ determination from a measurement of the muon forward-backward asymmetry at the FCC-ee, with a one-year running at any given centre-of-mass energy. The best accuracy is obtained for one year of running either just below or just above the Z pole, at 87.9 and 94.3\,GeV, respectively.}
\end{center}
\end{figure}

At a given centre-of-mass energy $\sqrt{s}$, the ${\rm e^+e^-} \to \mu^+\mu^-$ production cross section, $\sigma_{\mu\mu}$, is the sum of three terms: the photon-exchange term, ${\cal G}$, proportional to $\alpha^2_{\rm QED}(s)$; the Z-exchange term, ${\cal Z}$, proportional to $G_F^2$ (where $G_F$ is the Fermi constant); and the Z-photon interference term, ${\cal I}$, proportional to $\alpha_{\rm QED}(s) \times G_F$. The muon forward-backward asymmetry, $A_{\rm FB}^{\mu\mu}$, is maximally dependent on the interference term
\begin{equation}
\label{eq:AFBsimple}
A_{\rm FB}^{\mu\mu} = A_{\rm FB,0}^{\mu\mu} + \frac{3}{4} \frac{{\mathscr a^2}}{{\mathscr v^2}} \frac{\cal I}{{\cal G}+{\cal Z}}, 
\end{equation}
(where ${\mathscr a} = -0.5$, ${\mathscr v} = {\mathscr a} \times ( 1 - 4\sin^2\theta_{\rm W}) \simeq -0.037$, and $A_{\rm FB,0}^{\mu\mu} = 3{\mathscr v^2 a^2}/({\mathscr a^2+v^2})^2 \simeq 0.016$ is the small asymmetry at the Z pole), hence varies with $\alpha_{\rm QED}(s)$ as follows:
\begin{equation}
\Delta A_{\rm FB}^{\mu\mu} = \left( A_{\rm FB}^{\mu\mu} - A_{\rm FB,0}^{\mu\mu} \right) \times \frac{{\cal Z}-{\cal G}}{{\cal Z}+{\cal G}} \times \frac{\Delta\alpha}{\alpha}.
\label{eq:sensitivity}
\end{equation}
This expression shows that the asymmetry is not sensitive to $\alpha_{\rm QED}$ when the Z- and photon-exchange terms are equal, {\it i.e.}, at $\sqrt{s} = 78$ and $112$\,GeV, where the asymmetry is maximal. Similarly, the sensitivity to the electromagnetic coupling constant vanishes in the immediate vicinity of the Z pole. A maximum of sensitivity is therefore to be expected between $78$\,GeV and the Z pole, on the one hand, and between the Z pole and $112$\,GeV, on the other. With the luminosity targeted at the FCC-ee in one year data taking, the statistical precision expected on $\alpha_{\rm QED}(s)$ is displayed in Fig.~\ref{fig:sensitivity} and indeed exhibit two optimal centre-of-mass energies, $\sqrt{s_-} \simeq 87.9$\,GeV and $\sqrt{s_+} \simeq 94.3$\,GeV. With one year of data at either energy,  the expected relative precision on $\alpha_\pm \equiv \alpha_{\rm QED}(s_\pm)$ is of the order of $3\times 10^{-5}$, a factor four smaller than today's accuracy. 

The large uncertainty related to the running of the electromagnetic constant, inherent to the measurements made at low-energy colliders, totally vanishes at the FCC-ee with the weighted average of the two measurements of $\alpha_-$ and of $\alpha_+$:
\begin{equation}
\label{eq:xi}
\frac{1}{\alpha_0} = \frac{1}{2} \left( \frac{ 1 - \xi}{\alpha_-} + \frac{1+\xi}{\alpha_+} \right), {\rm \ \ \ where \ \ \ } \xi = \frac{\log s_-s_+ / m_{\rm Z}^4 }{\log s_- / s_+ } \simeq 0.045. 
\end{equation}
This combination has other advantages, the most important of which is the cancellation to a large extent of many systematic uncertainties, common to both measurements, but affecting $\alpha_-$ and $\alpha_+$ in opposite direction because of the change of sign of $A_{\rm FB}^{\mu\mu} - A_{\rm FB,0}^{\mu\mu}$ below and above the Z peak (Eq.~\ref{eq:sensitivity}).

A comprehensive list of sources for experimental, parametric, theoretical systematic uncertainties are examined in Ref.~\cite{Janot:2015gjr}. Most of these uncertainties are shown to be under control at the level of $10^{-5}$ or below, as summarized in Table~\ref{tab:summary1}, often because of the aforementioned delicate cancellation between the two asymmetry measurements. The knowledge of the beam energy, both on- and off-peak, turns out to be the dominant contribution, albeit still well below the targeted statistical power of the method.   
\begin{table}
\begin{center}
\begin{tabular}{|c|l|r|}
\hline Type & Source & Uncertainty \\ \hline\hline
                  & $E_{\rm beam}$ calibration & $1\times 10^{-5}$  \\
                  & $E_{\rm beam}$ spread      & $< 10^{-7}$  \\
Experimental      & Acceptance and efficiency  & negl.  \\
                  & Charge inversion & negl.  \\
                  & Backgrounds & negl. \\ \hline
                  & $m_{\rm Z}$ and $\Gamma_{\rm Z}$ & $1 \times 10^{-6}$ \\
Parametric        & $\sin^2\theta_{\rm W}$ & $5 \times 10^{-6}$ \\
                  & $G_{\rm F}$ & $5 \times 10^{-7}$ \\ \hline
                  & QED (ISR, FSR, IFI) & $< 10^{-6}$ \\ 
Theoretical       & Missing EW higher orders & few $10^{-4}$ \\ 
                  & New physics in the running &  $0.0$ \\ \hline\hline
Total            & Systematics & $1.2 \times 10^{-5}$ \\ 
(except missing EW higher orders)              & Statistics  & $3 \times 10^{-5}$ \\ \hline
\end{tabular}
\caption{Summary of relative statistical, experimental, parametric and theoretical uncertainties to the direct determination of the electromagnetic coupling constant at the FCC-ee, with a one-year running period equally shared between centre-of-mass energies of 87.9 and 94.3\,GeV, corresponding to an integrated luminosity of $85$\,${\rm ab}^{-1}$.}
\label{tab:summary1}
\end{center}
\end{table}

The fantastic integrated luminosity and the unique beam-energy determination are {\it the} key breakthrough advantages of the FCC-ee in the perspective of a precise determination of the electromagnetic coupling constant. 
Today, the only obstacle towards this measurement -- beside the construction of the collider and the delivery of the target luminosities -- stems from the lack of higher orders in the determination of the electroweak corrections to the forward-backward asymmetry prediction in the standard model. With the full one-loop calculation presently available for these corrections, a relative uncertainty on $A_{\rm FB}^{\mu\mu}$ of the order of a few $10^{-4}$ is estimated. An improvement deemed adequate to match the FCC-ee experimental precision might require a calculation beyond two loops, which may be beyond the current state of the art, but is possibly within reach on the time scale required by the FCC-ee. 

A consistent international programme for present and future young theorists must therefore be set up towards significant precision improvements in the prediction of all electroweak precision observables, in order to reap the rewards potentially offered by the FCC-ee. 

\section{QED interference effects in muon charge asymmetry near Z peak}
{\em by S. Jadach}

Experimenta data for $M_Z, G_F, \alpha_{QED}(0)$ are most important input
in the SM overall fit to experimental data.
However, $\alpha_{QED}(Q^2=0)$ is ported to $Q^2=M^2_{Z}$
using low energy hadronic data -- this limits its usefulness beyond LEP precision.
Patrick Janot has proposed \cite{Janot:2015gjr} another observable,
$A_{FB}(e^+e^-\to \mu^+\mu^-)$ at $\sqrt{s_\pm}=M_Z \pm 3.5 GeV$,
with a similar ''testing profile'' in the SM overall  fit as  $\alpha_{QED}(M^2_{Z})$ ,
but could be measured at high luminosity FCCee very precisely.
\footnote{It is advertised as
  ``determining  $\alpha_{QED}(M^2_{Z})$'' from $A_{FB}(\sqrt{s_{\pm}})$''.}
However, $A_{FB}$ near $\sqrt{s_{\pm}}$ is varying very strongly,
hence is prone to large QED corrections due to initial state bremsstrahlung.
Moreover, away from Z peak, it gets also 
a direct sizable contributions from QED initial-final state interference,
nicknamed at LEP era as IFI.
It is therefore necessary to re-discuss how efficiently
these trivial but large QED effects in $A_{FB}$
can be controlled and/or eliminated.

\begin{figure}[h!]
\begin{center}
\includegraphics[width=0.84\columnwidth]{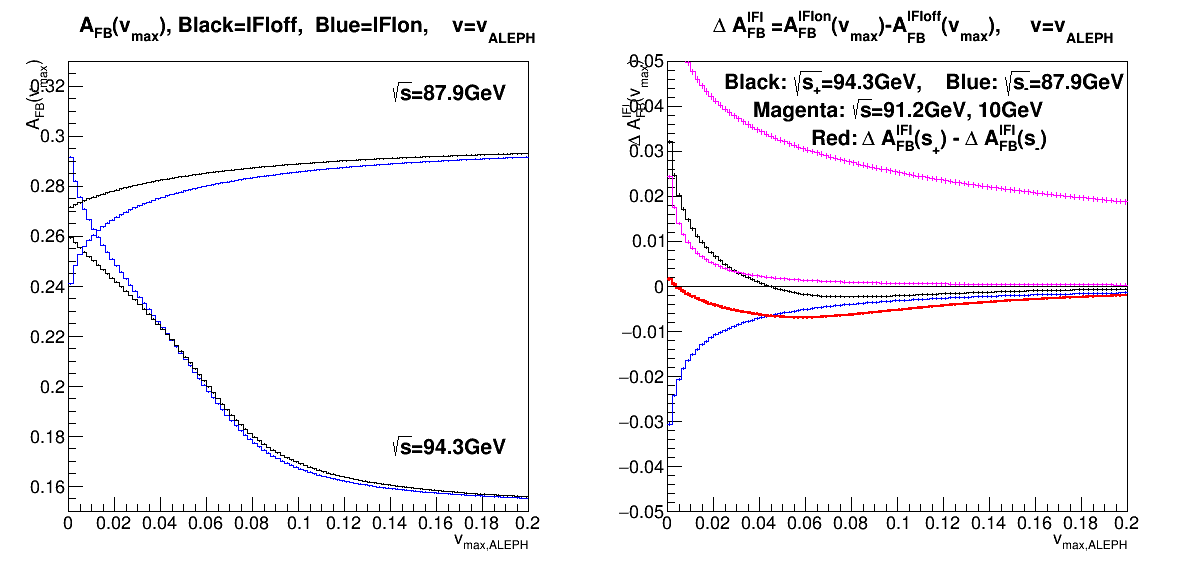}
\caption{\label{fig:IFI}
Direct influence of IFI in $A_{FB}(e^+e^-\to\mu^+\mu^-)$ 
at $\sqrt{s}\sim M_Z \pm 3$GeV.
Sign of $A_{FB}(87.9GeV)$ flipped in order to better fit into plot.
Numerical results are obtained using KKMC program \protect\cite{Jadach_2001}.
}
\end{center}
\end{figure}

Using numerical results from KKMC program \cite{Jadach_2001},  Fig.\ref{fig:IFI} 
illustrates  well known IFI suppression 
in AFB near resonance by factor $\sim\Gamma/M$,  when comparing
contribution of IFI at $\sqrt{s}=10$GeV and $91$GeV.
Variable $v_{\max}$ represents (dimensionless) upper limit on the total photon energy
in units of the beam energy.
The IFI effect is $\sim 3\%$ near $Z$ peak at $s_\pm$ (down to $\sim 1\%$ when combined).
Hence the effect of IFI is huge, 
compared to the aimed precision $\delta A_{FB} \sim 10^{-5}$.
Note that $\sim\Gamma/M$ suppression dies out for $v_{\max}<0.04$.
Let us also stressed,
that for stronger cut-off on photon emissions IFI effect in $A_{FB}$ 
is not smaller but bigger!
The above result was obtained using latest version of KKMC 4.22, which is available
on the FCC wiki page https://twiki.cern.ch/twiki/bin/view/FCC/Kkmc.

Is there any hope to control such a huge IFI effect extremely precisely?
The precedes of the $Z$ lineshape, where QED effects ia also huge, 30\%,
and finally has turned out to be controlable in QED calculation well below 0.1\%, 
shows that it may be possible.
Especially that in both cases, IFI in $A_{FB}$ and ISR in $Z$ lineshape,
both are due to well known physics of the multiple soft photon radiation.

\begin{figure}[h!]
\begin{center}
\includegraphics[width=0.7\columnwidth]{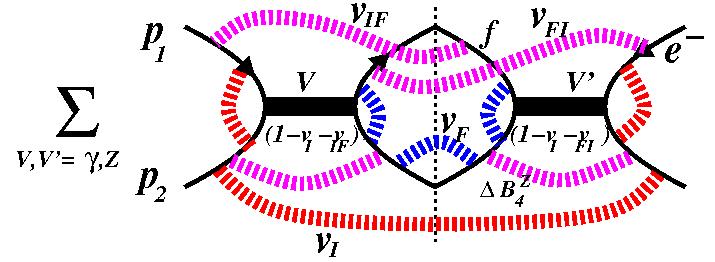}
\caption{\label{fig:IFIformula}
Graph illustrating Eq.~(12) with wide dashed lines representing multiphoton exchanges. Cut lines represent real photons while other ones depict exchange of virtual photon.
}
\end{center}
\end{figure}

Preliminary study using KKMC, in which
the difference between two versions of the IFI implementations was examined,
provides optimistic uncertainty estimate $\delta A_{FB} < 4\cdot 10^{-4}$ for $v_{\max}<0.03$,
that is comparable with the experimental goal.
However, this result should not be trusted, unless more systematic studies are done.
Generally, KKMC has the best, and possibly sufficient implementation of the IFI
effect taking multiphoton resummation.
However, one should validate KKMC predictions with a semianalytical calculation,
which was already proposed in ref. \cite{Jadach_2001},
extending earlier pioneering works of Frascati group \cite{Greco_1975}and \cite{Greco_1978},
but was not implemented in a numerical program.
It was not done, because it is technicaly a litle bit complicated --
instead of one-dimensional integration in the lineshape case, 
it involves 3-dimensional integral, to complicated to be done analytically.
At the LEP times it was also not urgent to do the above exercise, 
because of sizeable experimental errors of LEP data for $A_{FB}$.
Such a formula for the QED effects in the muon pair angular distribution,
based on the soft photon resummation \cite{Jadach_2001}
and depicted in Fig. \ref{fig:IFIformula},
has the following structure:
\begin{equation}
\begin{split}
\label{eq:mpg}
 {d\sigma \over d\Omega}(s,\theta,v_{\max}) & = 
        \sum_{V,V'=\gamma,Z}\;
        \int dv_I\;  dv_F\;  dv_{IF}\;  dv_{FI}\; 
        \delta(v-v_I-v_F-v_{IF}-v_{FI}) \theta(v<v_{\max})
        \\
& \times F(\gamma_I) \gamma_I v_I^{\gamma_I-1}\;
         F(\gamma_F) \gamma_I v_F^{\gamma_F-1}\;
         F(\gamma_{IF}) \gamma_{IF} v_{IF}^{\gamma_{IF}-1}\;
         F(\gamma_{FI}) \gamma_{FI} v_{IF}^{\gamma_{FI}-1}\;
         \\
& \times 
        e^{ 2\alpha  \Delta B_4^V}
        M^{(0)}_V\big(s(1-v_I-v_{IF}),\theta\big)\;
        [e^{ 2\alpha  \Delta B_4^{V'}}
        M^{(0)}_{V'}\big(s(1-v_I-v_{FI}),\theta\big)]^*\; 
        \big[1+\omega(v_I,v_F)\big],
\end{split}
\end{equation}
where $F(\gamma)\equiv \frac{e^{-C_E\gamma}}{\Gamma(1+\gamma)}$,
$\gamma_I =\frac{\alpha}{\pi} [ \frac{s}{m_e^2} -1]$,\quad
$\gamma_F =\frac{\alpha}{\pi} [ \frac{s}{m_\mu^2} -1]$,\quad
$\gamma_{IF} =\gamma_{FI}=\frac{\alpha}{\pi} \ln\frac{1-\cos\theta}{1+\cos\theta}$,\quad
the non-infrared remnant is $\omega(v_I,v_F)$ 
and $\Delta B_4^{V}$ is a resonance addition to virtual form-factor.
Once the above formula is implemented in a numerical form and compared with KKMC,
we shall gain solid estimate of the uncertainty of the QED corrections to $A_{FB}$
near $Z$ resonance.

\section{Precision measurements: sensitivity to new physics scenarios}
\label{sec:Erler}
{\it Jens Erler}

In the following, I will assume the very optimistic case, where the theory uncertainties from unknown higher orders will not be dominant.  
Progress has been steady in the past, and many types of radiative corrections can be included even after the FCC-ee era.
The fit results described below derive from the observable list in Table~\ref{EWobservables}.

\begin{table}[h]
\begin{center}
\begin{tabular}{|l|l|l|l|}
\hline
observable & current & FCC-ee & comment \\
\hline
$M_Z$ & $\pm 2.1$~MeV & $< 100$~keV & limited by systematics \\
$\Gamma_Z$ & $\pm 2.3$~MeV & $< 100$~keV & limited by systematics \\
$R_\mu$ & $\pm 0.025$ & $< 0.001$ & limited by systematics \\
$R_b$ & $\pm 6.6 \times 10^{-4}$ & $< 6 \times 10^{-5}$ & limited by systematics \\
$m_t$ & $\pm 810$~MeV & $\pm 15$~MeV & current error includes QCD uncertainty \\
$\sigma_{\rm had}^0$ & $\pm 37$~pb & $\pm 4$~pb & assumes 0.01\% luminosity error at FCC \\
$A_{LR}$ & $\pm 0.0022$ &$\pm 2 \times 10^{-5}$ & needs 4-loop calculation to match exp. \\
$A_{LR}^{FB}(b) $ & $\pm 0.0020$ & $\pm 0.0001$ & if no polarization at FCC use $A_{FB}(b)$  \\
$M_W$ & $\pm 15$~MeV & $\pm 0.6$~MeV & LEP precision was $\pm 33$~MeV \\
$\Gamma_W$ & $\pm 42$~MeV & $\pm 1$~MeV & 1st + 2nd row CKM unitarity test \\
$m_b$ & $\pm 23$~MeV & $\pm 9$~MeV & using $H \to b\bar b$ branching ratio at FCC \\
$m_c$ & $\pm 34$~MeV & $\pm 8$~MeV & using $H \to c\bar c$ branching ratio at FCC \\
$\Delta\alpha_{\rm had}$ & $\pm 1.3 \times 10^{-4}$ & $\pm 1.8 \times 10^{-5}$ & from $\sigma(\mu)$ and $A_{FB}(\mu)$ at FCC near $M_Z$ \\ 
\hline
\end{tabular}
\caption{\label{EWobservables}
Current and future uncertainties of key electroweak observables. The FCC-ee projections are taken from or motivated by the target uncertainties in Ref.~\cite{Gomez-Ceballos:2013zzn}.}
\end{center}
\end{table}

If one succeeds to extract these (pseudo)-observables with the indicated precision,
a fit to the SM parameters would result in a determination of
\begin{eqnarray}
\nonumber M_H \mbox{ within }  \pm 1.3 \mbox{ GeV} & 
\mbox{currently: } M_H = 96^{+22}_{-19} \mbox{ GeV} \\ 
\nonumber \alpha_s (M_Z) 
\mbox{ within }  \pm 0.00009 & \hspace{45pt}
\mbox{currently: } \alpha_s (M_Z) =  0.1181 \pm 0.0013  
\end{eqnarray}
where the $\alpha_s$ projection includes further non-$Z$-pole determinations that would be possible at the FCC \cite{Agashe:2014kda},
such as from $W$ decays, deep inelastic scattering and jet-event shapes.
Likewise, the number of active neutrinos $N_\nu$ can be constrained within $\pm 0.0006$ 
compared to the current result $N_\nu = 2.992 \pm 0.007$.

\begin{figure}[h!]
\begin{center}
\includegraphics[width=0.70\columnwidth]{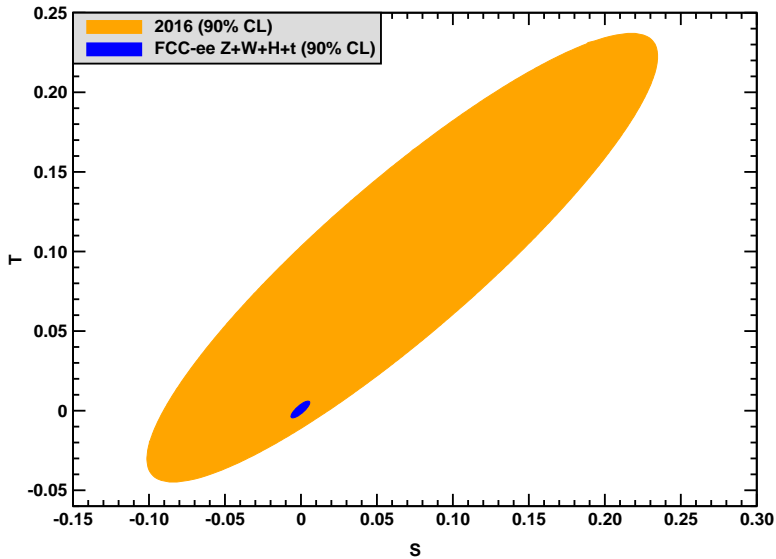}
\caption{\label{ST}
Current and FCC-ee constraints on the oblique parameters $S$ and $T$.}
\end{center}
\end{figure}

An important benchmark are the {\em oblique parameters\/} (see Fig.~\ref{ST}) describing new physics contributions 
to the gauge boson self-energies.
{\em E.g.}, $T$ (or $\rho_0$) would constrain VEVs of higher dimensional Higgs representations within $\lesssim 1$~GeV, 
while {\em degenerate\/} scalar doublets could be probed up to 2~TeV~\cite{Henning:2014wua}.
Assuming that the FCC-ee would not see a deviation from the SM ($\rho_0 = 1 \pm 0.00012$),
or alternatively that the central value would be unchanged from today ($\rho_0 = 1.00037 \pm 0.00012$), 
the mass splittings of non-degenerate doublets of heavy fermions would satisfy
($C_i$ is the color factor),
$$
\sum_i {C_i \over 3} \Delta m_i^2 \leq (8 \mbox{ GeV})^2 \hspace{45pt}
\sum_i {C_i \over 3} \Delta m_i^2 = (34 \pm 1 \mbox{ GeV})^2,
$$
respectively, compared to the current limit of $(49 \mbox{ GeV})^2$.
Model-independently, $T$ would be sensitive to new physics with $\cal O$(1) couplings 
up to scales of about 70~TeV.
Also, the precision in the parameters entering the $Zbb$-vertex 
(one of which currently showing a 2.7~$\sigma$ deviation) would improve by an order of magnitude.

\subsubsection*{Acknowledgements}
I gratefully acknowledge support by Alain Blondel and the University of Geneva.

\section{Composite Higgs Models at $e^+ e^-$ Colliders}
\label{sec:DeCurtis}
{\it Stefania De Curtis} 

A future $e^+e^-$ collider will be capable to show the imprint of  composite Higgs scenarios encompassing partial compositeness. Besides the detailed study of the Higgs properties, such a machine will have a rich top-quark physics program mainly in two domains: top property accurate determination at the $t \bar t$ production threshold, search for New Physics  (NP) with top quarks above the threshold. In both domains, a composite Higgs scenario can show up. Here we discuss such possibility using a particular realisation, namely the 4-Dimensional Composite Higgs Model (4DCHM) \cite{DeCurtis:2011yx}. It describes the intriguing possibility that the Higgs particle may be a composite state arising from some strongly interacting dynamics at a high scale. This will solve the hierarchy problem owing to compositeness form factors taming the  divergent growth of the Higgs mass upon quantum effects. Furthermore, the now measured light mass could be  consistent with the fact that the composite Higgs  arises as a pseudo Nambu-Goldstone Boson (pNGB) from a particular coset of a global symmetry breaking.
Models with a pNGB  Higgs generally predict modifications of its couplings to both bosons and fermions of the SM, hence the measurement of these quantities represents a powerful way to test its possible non-fundamental nature. Furthermore, the presence of additional particles in the spectrum of such composite Higgs models (CHMs) leads to both mixing effects with the SM states as well as new Feynman diagram topologies both of which would represent a further source of deviations from the SM expectations.

In the near future, the LHC will be able to test Beyond-Standard-Model (BSM) scenarios  more extensively, probing the existence of new particles predicted by extensions of the SM to an unprecedented level. Nevertheless the expected bounds, though severe, could not be conclusive to completely exclude natural scenarios for the Fermi scale. As an example,  new gauge bosons predicted by  CHMs with mass larger than $\sim$ 2~TeV could escape the detection of the LHC \cite{Barducci:2012kk}.  
Furthermore, concerning the Higgs properties, the LHC will not be able to measure the Higgs couplings  to better than  few \%  leaving room to scenarios, like CHMs, which predict deviations within the foreseen experimental accuracy for natural choices of the compositeness scale $f$, (namely in the TeV range) and of the strong coupling constant $g_\rho$.  For these reasons  we  will face here the case in which LHC will not discover a $W'/Z'$ (or not be able to clearly asses its properties) and also will not discover any extra fermion (or it will discover it with a mass around 800 GeV, which is roughly the present bound, but without any other hint about the theory to which it belongs).  In this situation, an $e^+e^-$ collider could have a great power for enlightening indirect effects of  BSM physics.

\begin{figure}[h!]
\begin{center}
\includegraphics[width=0.80\linewidth]{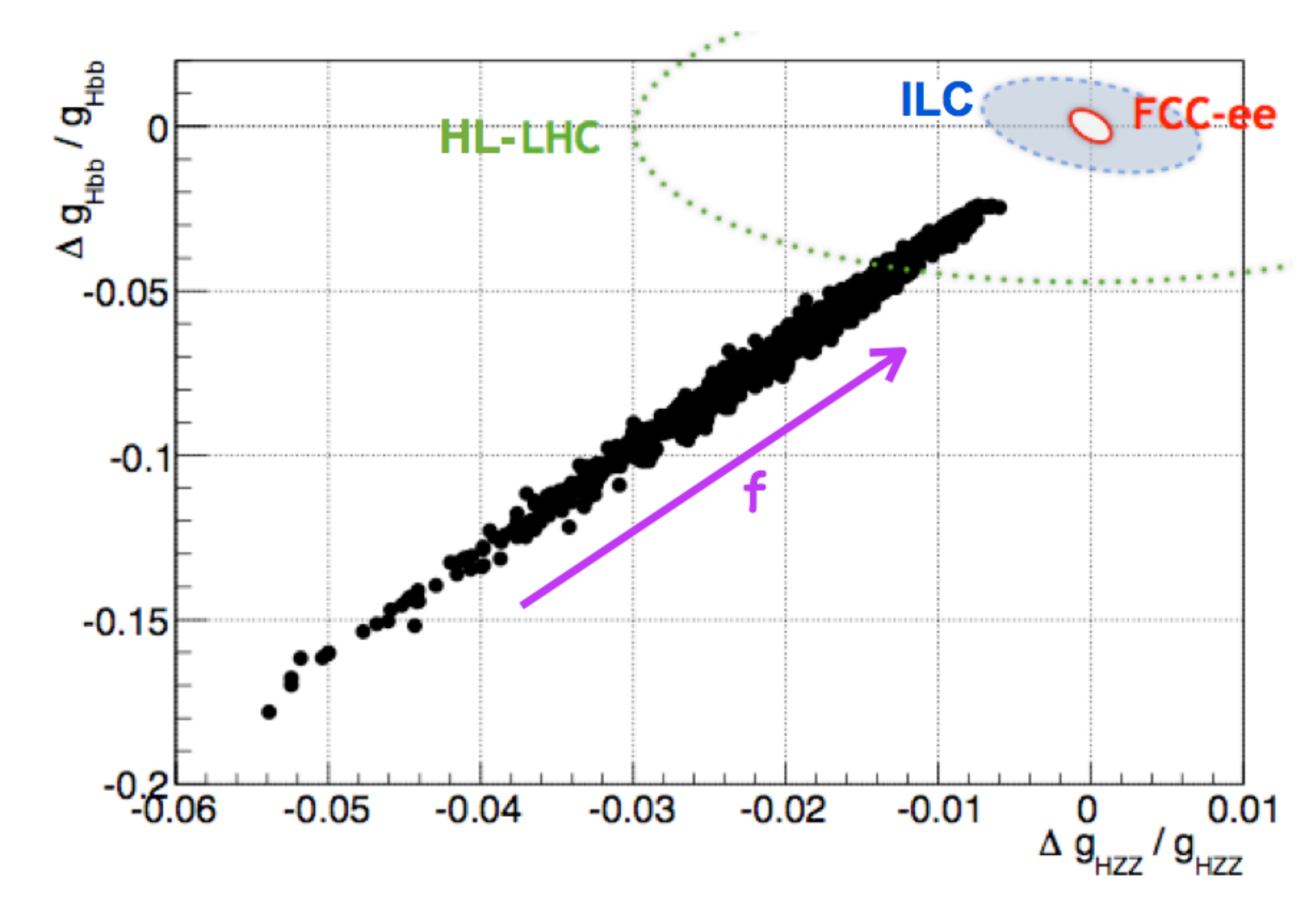}
\caption{\label{fig1}
Deviations for the $HZZ$ and $Hb\bar b$ couplings in the 4DCHM (black points) compared with the relative precision expected at HL-LHC, ILC, FCC-ee \protect\cite{Janot:2015mqv}. 
}
\end{center}
\end{figure}

\begin{figure}[h!]
\begin{center}
\includegraphics[width=0.45\linewidth]{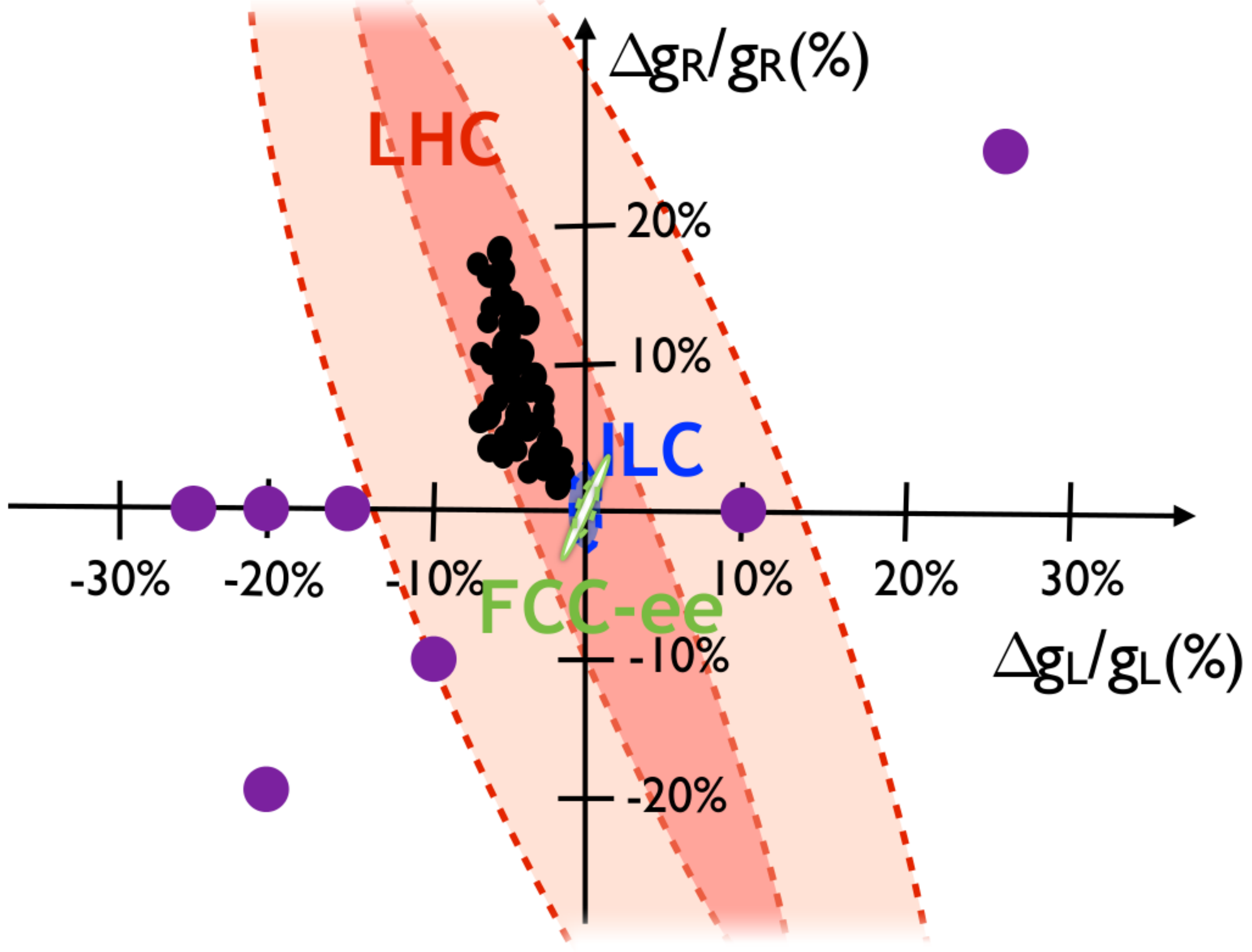}
\includegraphics[width=0.40\linewidth]{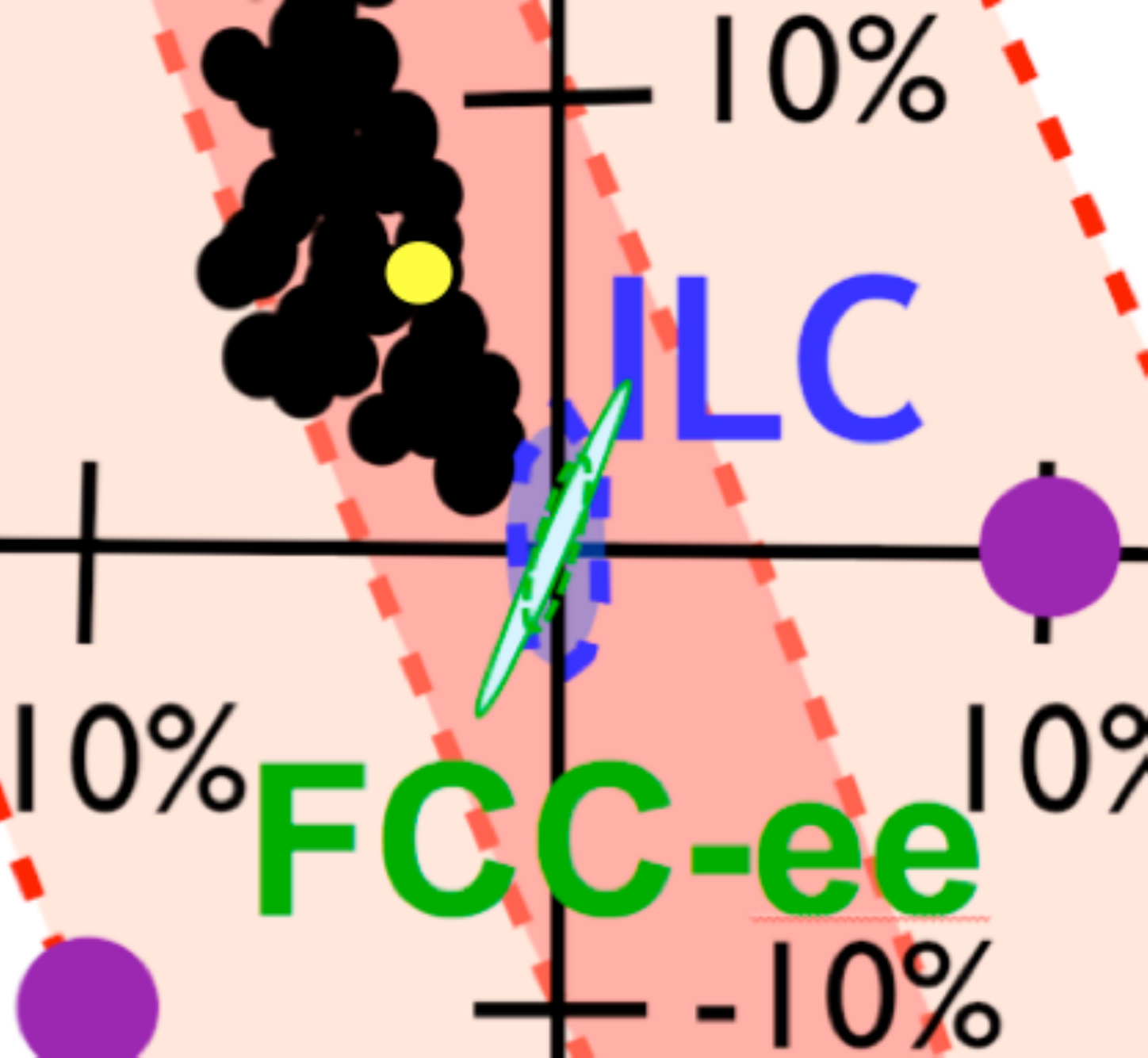}
\caption{
Left: Typical deviations for the $Zt_L \bar t_L$ and $Z t_R \bar t_R$ couplings in various NP models represented by purple points (see~\protect\cite{Richard:2014upa}) and in the 4DCHM (black points). Also shown are the sensitivities expected after LHC-13 with 300 fb$^{-1}$, (region inside the red-dashed lines), after HL-LHC with 3000 fb$^{-1}$ (region inside the inner red-dashed lines), from ILC-500 with polarised beams  (region inside the blue-dashed lines) and from FCC-ee (region inside the green lines: the continuous(dashed) line indicates the bounds extracted from the angular and energy distribution of leptons($b$-quarks))~\protect\cite{Barducci:2015aoa,Janot:2015yza,Andreazza:2015bja}.
The 4DCHM black points correspond to a scan with $0.75<f({\rm TeV})<1.5$, $1.5<g_\rho<3$  and on the extra-fermion sector parameters as described in \protect\cite{Barducci:2012kk} with  the constraints: $M_{Z'}\sim f g_\rho>2$  TeV and $M_{T}>800$ GeV with $T$ the lightest extra-fermion; Right: Deviations in the $Zt_L \bar t_L$ and $Z t_R \bar t_R$ couplings for the 4DCHM benchmark point-A (yellow point)
\label{fig2a}%
}
\end{center}
\end{figure}

The main Higgs production channels within the 4DCHM were considered in \cite{Barducci:2013ioa} for three possible energy stages and different luminosity options of the proposed $e^+e^-$ machines, and the results were confronted to the expected experimental accuracies in the various Higgs decay channels. Moreover the potentialities of  such colliders in discovering the  imprint of  partial compositeness in the top-quark sector through  an accurate determination of the top properties at the $t \bar t$ production threshold were analized in \cite{Janot:2015mqv}. In Fig.~\ref{fig1} we compare the deviations for the $HZZ$ and $Hbb$ couplings and in Fig.~\ref{fig2a}(left panel) for the $Z t_L\bar t_L$ and  $Z t_R\bar t_R$ couplings in the 4DCHM with the relative precision expected at HL-LHC, ILC, FCC-ee \cite{Janot:2015mqv,Barducci:2015aoa,Janot:2015yza,Andreazza:2015bja} 

From the Higgs coupling measurements, it is clear that the FCC-ee will be able to discover CHMs with a very large significance, also for values of $f$ larger than 1 TeV for which the LHC measurements will not be sufficient to detect deviations from the SM.  This is true {\it a fortiori} for the top electroweak coupling measurements. In fact an $e^+e^-$ collider can separately extract the left- and right-handed electroweak couplings of the top. This is particularly relevant for models with partially composite top, like the 4DCHM where the $Zt\bar t$ coupling modification comes not only from the mixing of the $Z$ with the $Z'$s but also from the mixing  of the top (antitop) with the extra-fermions (antifermions), as expected by the partial compositeness mechanism. For the "natural" scan described in the caption of Fig.~\ref{fig1}, the typical deviations lie within the region uncovered by  the HL-LHC but are well inside the reach of the polarized ILC-500 and the FCC-ee where the lack of initial polarization is compensated by the presence of a substantial final state polarization and by a larger integrated luminosity \cite{Barducci:2015aoa,Janot:2015yza,Andreazza:2015bja}.

But this is not the end of the story, in fact in CHMs the  modifications in the  $e^+e^-\to H Z $  and $e^+e^-\to t \bar t$ processes arise not only via the modification of the $HZZ$ coupling for the former and of the $Z t \bar t$ coupling for the latter,  but also from  the exchange of new particles, namely the $s$-channel exchange of t$Z'$s, which can be sizeable also for large $Z'$ masses due to the interference with the SM states.  This effect  can be crucial at high c.o.m energies of the collider but also important at moderate $\sqrt{s}$ \cite{Barducci:2015aoa}.  In particular it is impressive how the FCC-ee with $\sqrt{s}=365$ GeV and 2.6 ab$^{-1}$ (corresponding to 3 years of operation) could discover the imprint of  extra $Z'$ particles through their effective contribution to the EW top coupling deviations. This result emerges from the optimal-observable analysis of the lepton angular and energy distributions from top-quark pair production with semi-leptonic decays \cite{Grzadkowski:2000nx,Janot:2015yza,Janot:2015mqv}.
The $Vt\bar t$, $V=Z,\gamma$  vertices can be expressed in the usual way in terms of 8 form factors: 
\begin{eqnarray}
\Gamma_{Vt \bar t}^\mu=\frac g 2 \bar u(p_t)[\gamma^\mu\{A_V+\delta A_V-(B_V+\delta B_V)\gamma_5\}+\frac{(p_t-p_{\bar t})^\mu}{2 m_t}(\delta C_V- \delta D_V \gamma_5)]v(p_{\bar t})
\label{asym}
\end{eqnarray}
and the differential cross-section for the process:  $e^+e^- \to t \bar t\to (bW^+)(\bar b W^-)\to(bq q')(\bar b l \nu)$ can be expanded around their SM values:
\begin{eqnarray}
\frac{d^2 \sigma}{dx d\cos\theta}\sim S^0(x,\theta)+\sum_{i=1}^8 \delta_i f^i(x,\cos\theta),  ~~~ \delta_i=\delta(A,B,C,D)_V,~~~f^i=f^{A,B,C,D}_V(x,\cos\theta)
\label{xsec}
\end{eqnarray}
Here $S^0$ gives the SM contribution, $x$ and $\theta$ are the lepton reduced energy and the polar angle.
By considering only the 6 CP-conserving form factors ($A_V,B_V,C_V$), the elements of the covariance matrix (the statistical uncertainties)  are derived from a likelihood fit to the lepton angular/energy distributions and the total event rate \cite{Janot:2015yza}. The result for the top pair left- and right-handed couplings to the $Z$ is represented by the continous green ellipses in Fig. \ref{fig1}. 

In order to compare these uncertainties with the deviations expected in CHMs (deviations in the form factors due not only to coupling modifications but also to $Z'$s exchanges) we have considered one representative benchmark point of the 4DCHM (point-A) corresponding to $f=1.3$ TeV, $g_\rho$=1.5, $M$=1.4 TeV ($M$ is the scale of the extra-fermion mass). This scenario describes two nearly degenerate  $Z'$s, with mass $\sim$ 2.1 and 2.2 TeV respectively, which are active in the top pair production.  The deviations in the $Zt_L \bar t_L$ and $Z t_R \bar t_R$ couplings are $\delta g^Z_L/g^Z_L=-2.8\%$ and 
$\delta g^Z_R/g^Z_R=6.2\%$ as shown by the yellow point in Fig.~\ref{fig2a}(right) , while $\delta g^\gamma_L/g^\gamma_L=\delta g^\gamma_R/g^\gamma_R=0$.

\begin{figure}[h!]
\begin{center}
\includegraphics[width=0.45\linewidth]{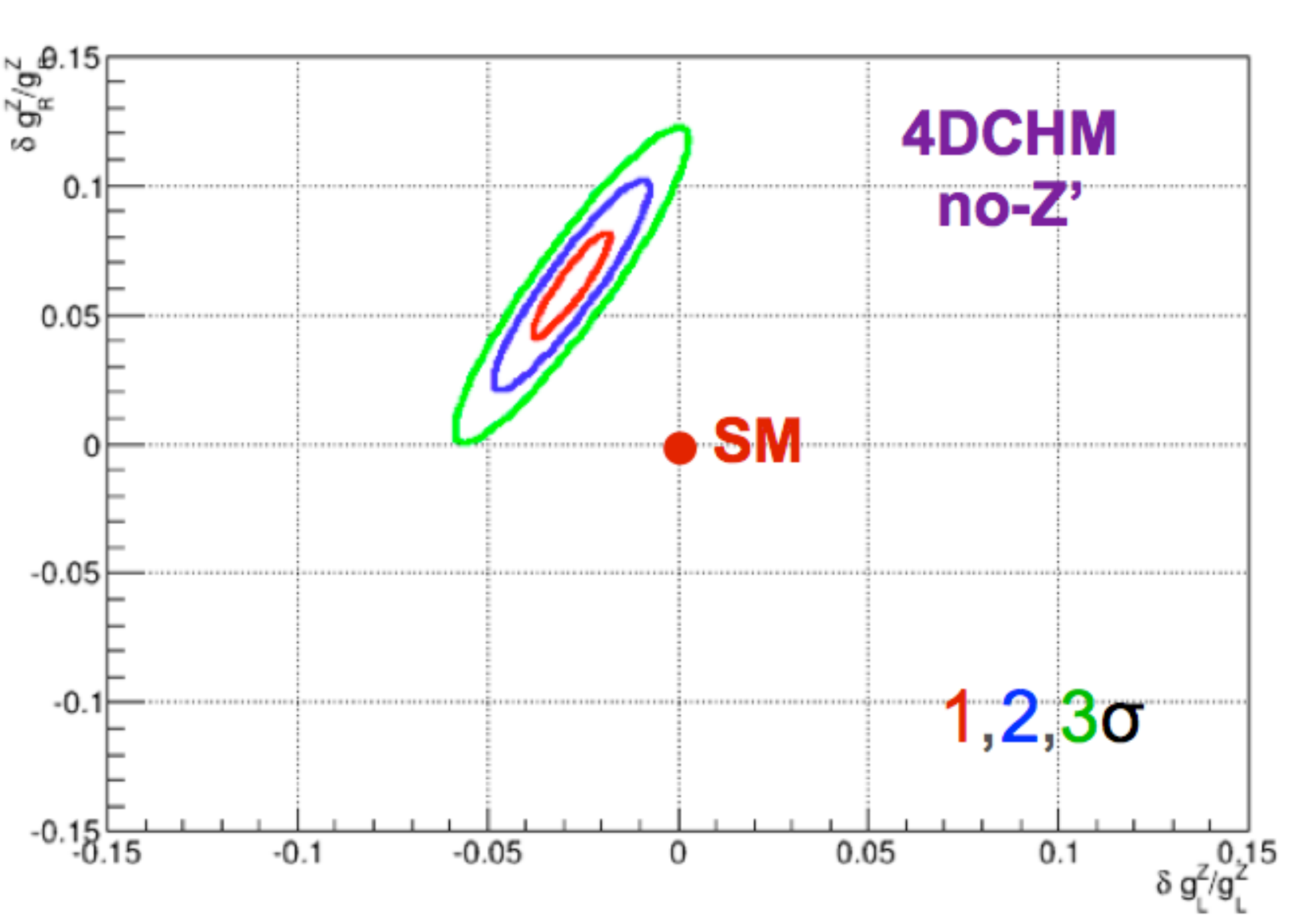}
\includegraphics[width=0.45\linewidth]{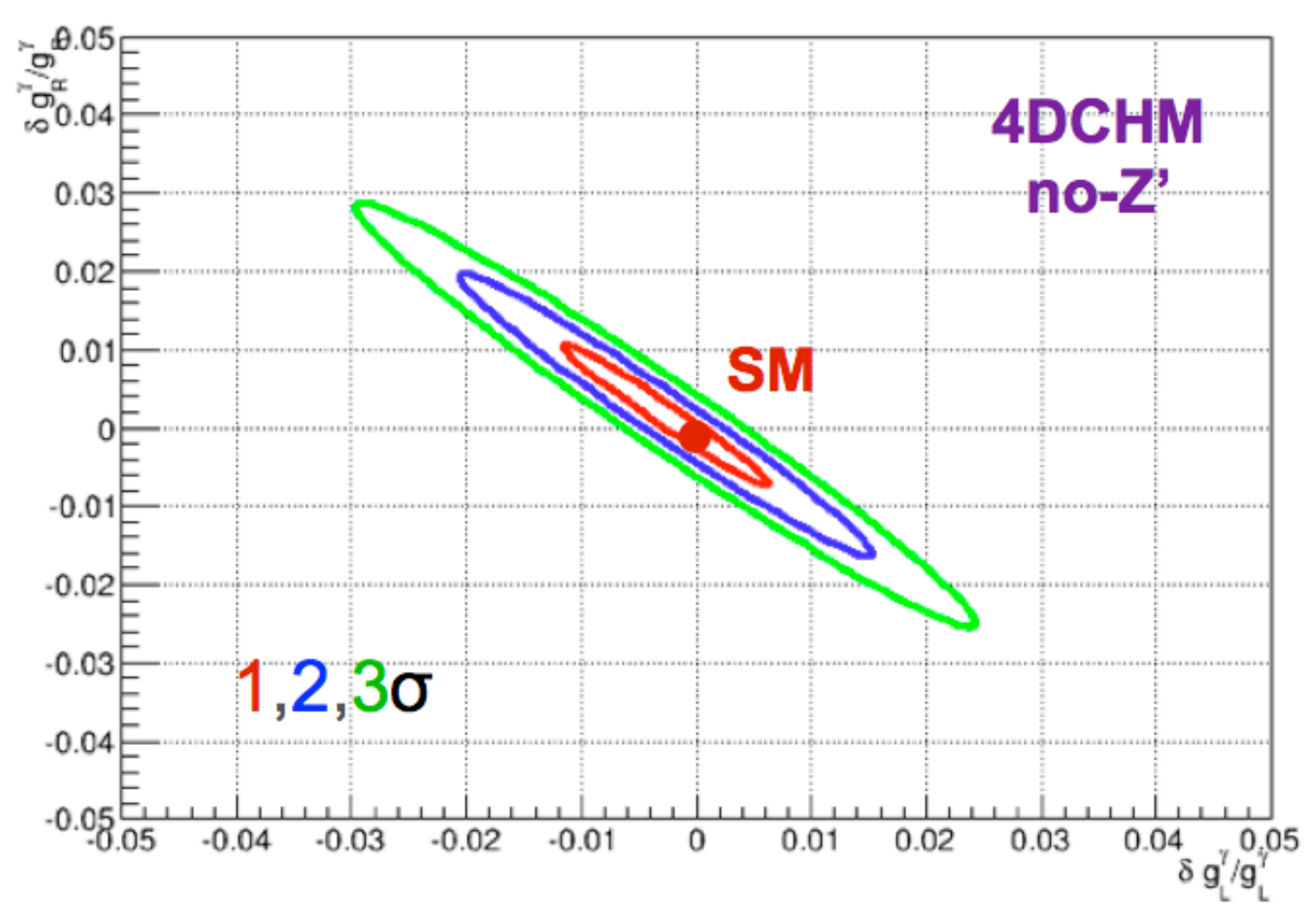}
\caption{\label{fig2b}
Left: Determination of the  $Zt_L \bar t_L$ and $Z t_R \bar t_R$ couplings from a 4 parameter fit of the optimal observable analysis for the point-A  without the $Z'$ exchanges. Right: Determination of the   $\gamma t_L \bar t_L$ and $\gamma  t_R \bar t_R$ couplings  from a 4 parameter fit of the optimal observable analysis for the point-A  without the $Z'$ exchanges.%
}
\end{center}
\end{figure}

As an exercise, we have evaluated  the  double differential cross section of Eq.~\ref{xsec} within the 4DCHM without including the $Z'$  exchanges and extracted the  deviations in the $Z$ and photon left and right couplings to the top pair by performing a 4 parameter fit (i.e. fixing the other two CP conserving form factors, $C_V$, to their SM value). The result is shown in Fig.~\ref{fig2b}. The ellipses correspond to 1,2,3 $\sigma$. As expected the deviations in the photon couplings are fully compatible with zero. Also, the central values of the $Z$ coupling deviations reproduce very well the deviations corresponding to the point-A ($\delta g^Z_{L(R)} {}^{\rm {point-A}}=-0.00713(-0.00708)$ to be compared with $\delta g^Z_{L(R)}{}^{\rm {Fit}}=-0.00717\pm 0.00475(-0.00701\pm0.00358)$ and $\delta g^\gamma_{L(R)} {}^{\rm {point-A}}=0$ to be compared with $\delta g^\gamma_{L(R)}{}^{\rm {Fit}}=-0.00056\pm 0.00224(0.00035\pm 0.00201)$ where we have included the marginalized uncertainties). The fit is very good in reproducing  the theoretical deviations for the couplings, but, as said, this is an exercise. In fact the $Z'$s were taken off. 

\begin{figure}[h!]
\begin{center}
\includegraphics[width=0.45\columnwidth]{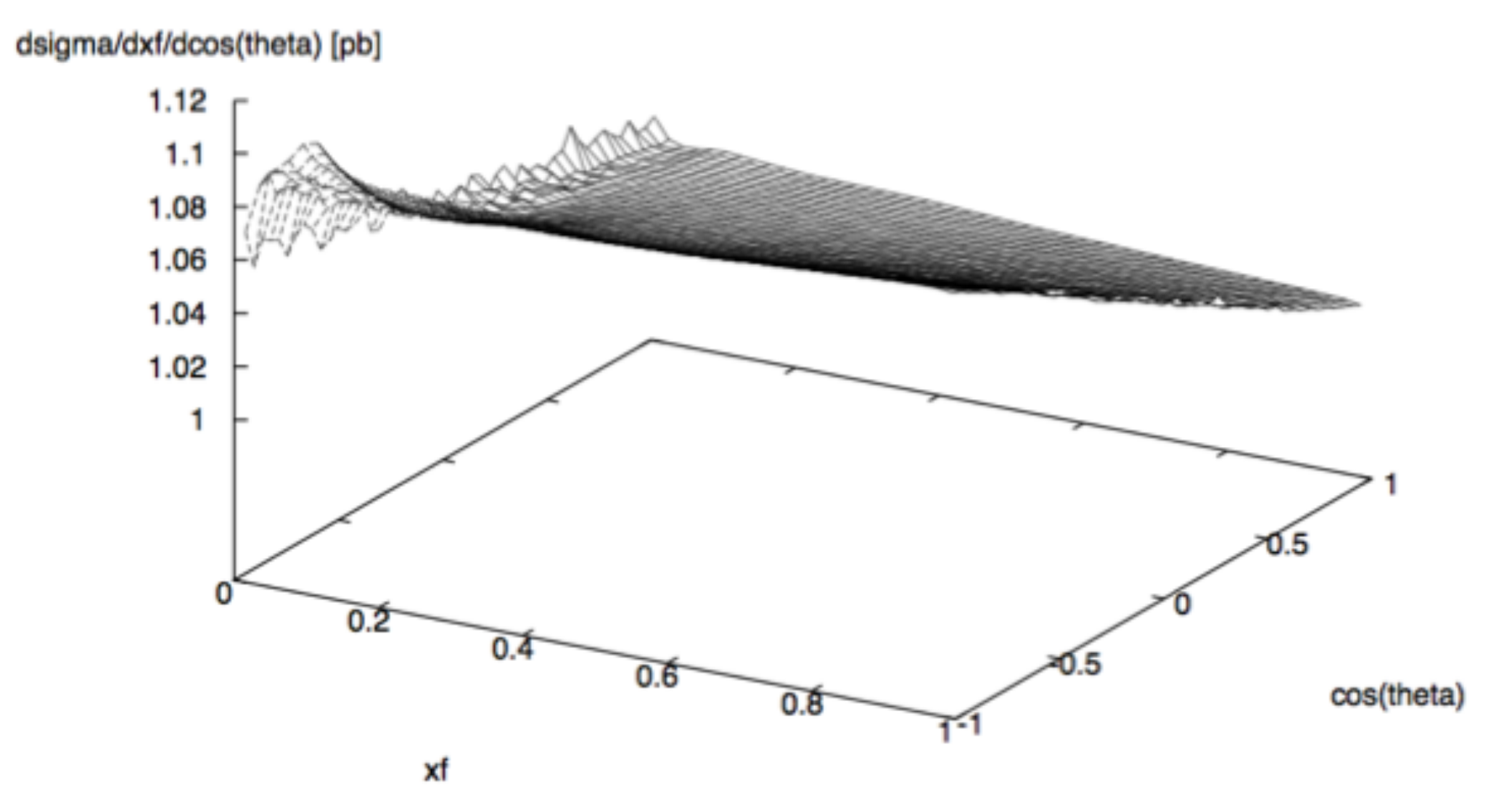}
\caption{\label{fig2d}
Double differential cross section with respect to the reduced lepton energy and the lepton polar angle within the 4DCHM normalized to the SM one.%
}
\end{center}
\end{figure}

\begin{figure}[h!]
\begin{center}
\includegraphics[width=0.45\linewidth]{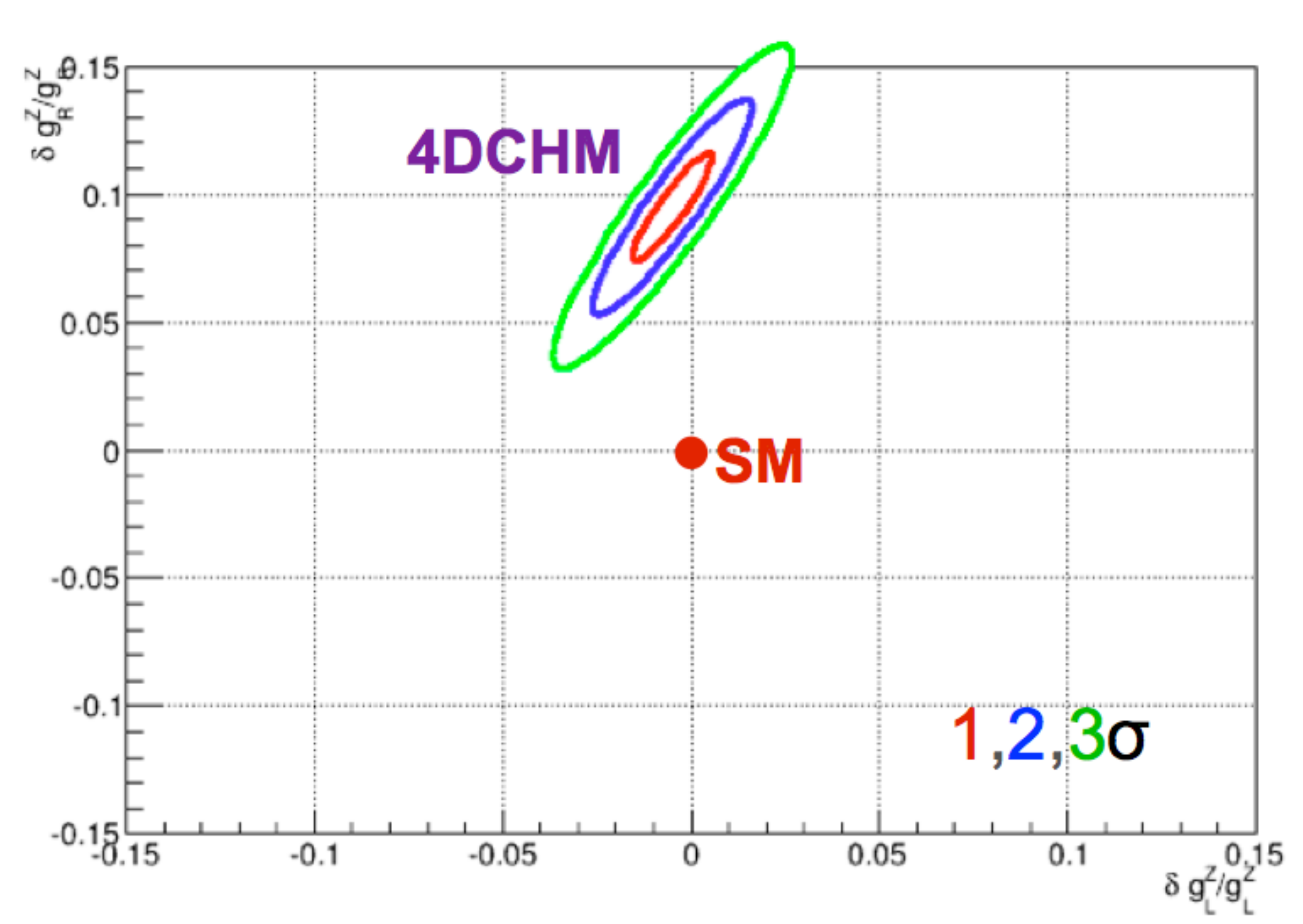}
\includegraphics[width=0.45\linewidth]{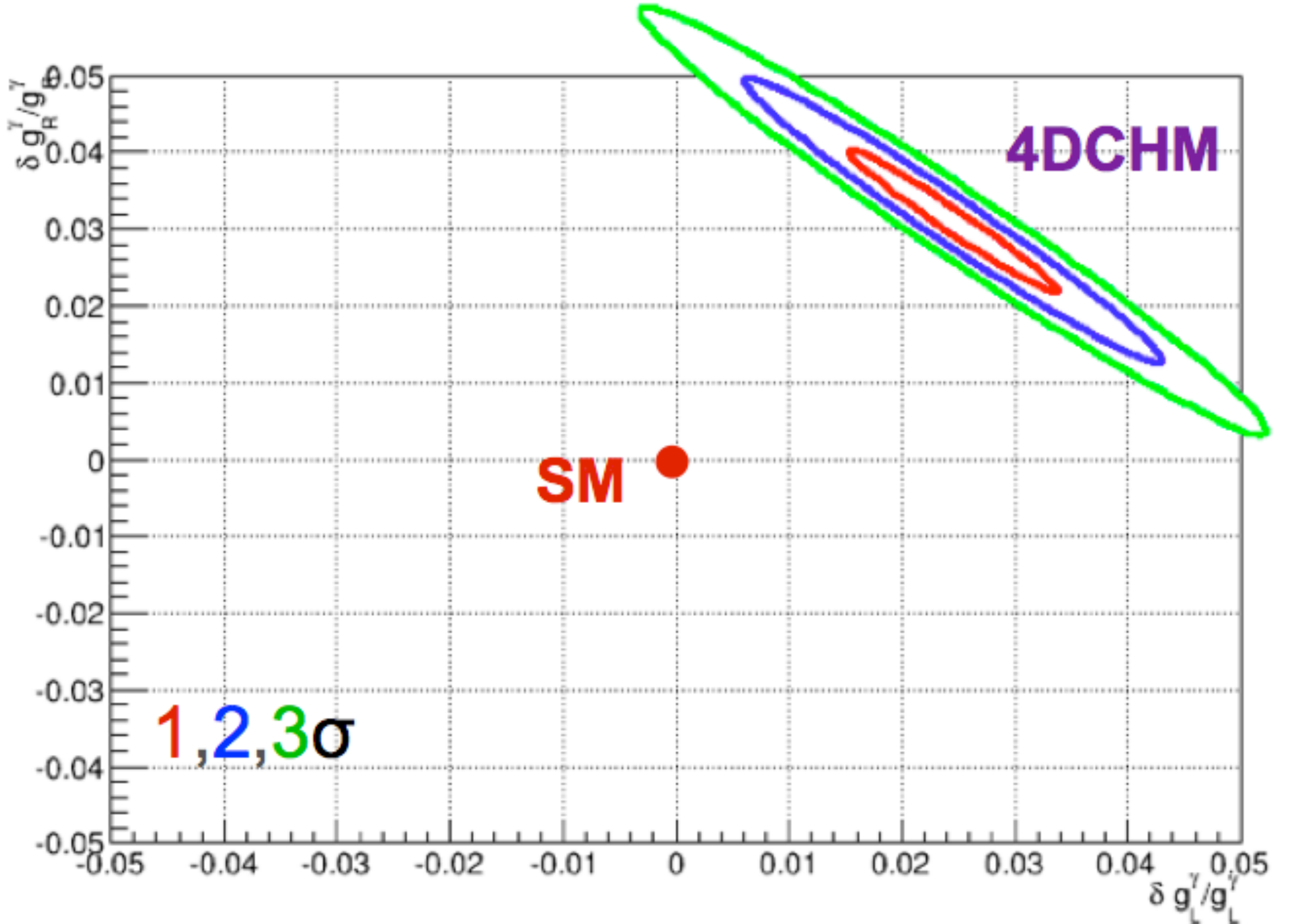}
\caption{\label{fig2e}
Left: Determination of the  $Zt_L \bar t_L$ and $Z t_R \bar t_R$ couplings from a 4 parameter fit of the optimal observable analysis for the full 4DCHM point-A. Right: Determination of the $\gamma t_L \bar t_L$  and $\gamma  t_R \bar t_R$  couplings from a 4 parameter fit of the optimal observable analysis for the full 4DCHM point-A%
}
\end{center}
\end{figure}

The double differential cross section within the full 4DCHM normalized to the SM one is shown in the bottom-left panel of Fig.~\ref{fig2d}, while the  deviations in the $Z$ and photon left and right "effective" couplings to the top pair extracted by performing a 4 parameter fit, are shown in Fig.~\ref{fig2e}. We called them "effective" couplings  because they include the effects of the interference of the $Z'$s with the SM gauge bosons.  This is clearly evident   by looking at  the deviations of the photon couplings,  which are completely due to these interference effects.

These results, derived here for a single benchmark point, are very promising and show the  importance of the interference between the SM and the $Z'$s  already at $\sqrt{s}$=365 GeV.  A more detailed analysis is worth to be done \cite{prep} to extract the dependence of the effective EW top couplings from the $Z'$ properties (mass and couplings, the width is not important at such moderate energies)  in order to clearly relate them to these new spin-1 particles which are naturally present in CHMs. 
The optimal-observable statistical analysis at FCC-ee offers a unique possibility to disentangle the effects of top coupling modifications (always taken into account in NP searches) from $Z'$ interference effects (often neglected).

\section{Constraints on top FCNC's from (hadronic) single top production in  $e^+e^-$ collisions}
\label{sec:Mele}
{\it  S.~Biswas,F.~Margaroli, B.~Mele}

Single top production in $e^+e^-$ colliders is a sensitive probe of anomalous 
Flavor Changing Neutral Currents (FCNC) in the top sector.
In particular, the $t\bar q \; (\bar t q)$ production (with $q=c,u$),  mediated by an $s$-channel photon and 
$Z$ vector boson has undetectably small rates in the standard model (SM), but 
might become visible for $tq\gamma$- and $tqZ$-coupling strenghts allowed by various beyond-the-SM theoretical frameworks, and not excluded by present experiments~\cite{Agashe:2013hma}. Single top production has two advantages with respect to top pair production above the $t\bar t$ threshold, where  FCNC's can be probed in top decays.
On the one hand, it can be studied at lower (more easily accessible) c.m. energies,
such as  $\sqrt s\simeq 240$~GeV which optimizes Higgs-boson studies.
On the other hand, since the FCNC $tq\gamma$ and $tqZ$ couplings in 
$e^+e^-\to t q$ are active in production, one can enhance the sensitivity to the magnetic-dipole-moment part ({\small $\sim q_{\gamma,Z}^\mu \,\sigma_{\mu\nu}$}) 
of the FCNC Lagrangian  by increasing
the collision energy $\sqrt s=${\small $\sqrt{q_{\gamma,Z}^2}$}. This feature can even make the single top production more sensitive to FCNC couplings than top-pair production at energies above the $t\bar t$ threshold~\cite{Agashe:2013hma}.

Here we report the results of a recent analysis of the $e^+e^-\to t q$ sensitivity reach on 
FCNC $tq\gamma$ and $tqZ$ couplings at $\sqrt s\simeq 240$~GreV, with 
the top quark decaying hadronically. Indeed, the exceptionally clean environment and the
great particle-identification capabilities expected at future $e^+e^-$ detectors will allow to fully exploit events with hadronic top decays, with a control and separation of background
never experienced before in top quark studies. The relevant FCNC couplings
(assuming no related parity violation) are defined by the lagrangian~:
{\scriptsize $
-\mathcal{L}^\mathrm{eff}  =  
e \lambda_{qt} \, \bar q 
\frac{i \sigma_{\mu \nu} q^\nu}{m_t} t A^\mu
+ \frac{g}{2 c_W} \mathcal{X}_{qt} \, 
\bar q \gamma_\mu  t Z^\mu 
+ \frac{g}{2 c_W} \kappa_{qt} \, \bar q 
\frac{i \sigma_{\mu \nu} q^\nu}{m_t} t Z^\mu   
+ \mathrm{h.c.} \, $}.
The {\scriptsize $\lambda_{qt},\mathcal{X}_{qt},\kappa_{qt}$} couplings can be straightforwardly related to the corresponding decay branching ratios 
{\scriptsize $\mathrm{BR}(t \to q \gamma) \simeq 0.43 \, \lambda_{qt}^2$,
 $ \mathrm{BR}(t \to qZ)_{\gamma_{\mu}} \simeq 0.47 \, \mathcal{X}_{qt}^2$,
 $\mathrm{BR}(t \to qZ)_{\sigma_{\mu,\nu}} \simeq 0.37 \, \kappa_{qt}^2$}.

 The 
{\small $e^+ e^- \rightarrow t j + \bar t j \rightarrow b W j \rightarrow b j j j$} signal,
and main  background  
{\small $e^+ e^- \rightarrow W j j \rightarrow j j j j$}\,
(other 4-jets background like {\small $q\bar{q}b\bar{b}$} have been checked to 
 be negligible) have been 
generated via MadGraph5\_aMC@NLO (with FeynRules modeling the FCNC interactions),  
and interfaced with PYTHIA for showering
and hadronization.
  Jets have been defined by PYTHIA iterative cone algorithm   with 
cone size {\small $R=0.4$}, and 
  jet energy resolution has been parametrized as {\small $\frac{\sigma(E)}{E}=30\%/\sqrt{E}$}. 
 The $b$-tagging efficiency and corresponding fake jet rejection for $c$ and light-quark jets are  crucial, since the background does not contain  $b$-quark initiated jets. 
  We have worked with two hypothesis : a) true $b$-jet tagging efficiency of {$\epsilon_b=60\%$}, 
 and corresponding $c$-(light)-jet rejection factor of {250 (1000)}; b)
  {$\epsilon_b=80\%$} with $c$-(light)-jet rejection factor 
 {10 (100)}~\cite{btagging}.
An optimized choice  of 
fake jet rejection factors indeed 
proved to  be more useful than a large
$b$-tagging efficiency.
 Basic cuts applied on jets are {\small  $p^{j,b}_T> 20$ GeV, $|\eta^{j,b}|<2.5$, {\rm and} $\Delta R(jj,bb,bj)>0.4$}.
 Through an MVA study we have then identified a further set of cuts, improving the background  rejection: a)  a di-jet system inside  
the mass window  {\small 65 GeV $<M_{jj}<$ 90 GeV}  (compatibility with $M_W$);
b) rejection of  events if invariant mass of the remaining di-jet system peaks around a second 
$W$ within a mass window  {\small 65 GeV $<M_{jj}<$ 85 GeV}; 
c) a jet  tagged as a $b$-jet is then combined with the reconstructed $W$, and required to  
produce a top system satisfying  {\small150 GeV $<M_{bW}<$ 175 GeV}; 
 d) finally, the jet which is neither a $b$-jet nor a jet coming from a $W$-decay is required 
    to have  {\small$|\vec{p}_{j}|<$65 GeV}.
    
 The final bounds on FCNC couplings and corresponding Br's  are reported for different integrated luminosities L in Table 1. With  L=100 fb$^{-1}$, 
 we find that the $e^+e^-\to t q$ hadronic top channel  
 has about twice the sensitivity to BR's as the semileptonic  top
channel~\cite{Khanpour:2014xla}.



\begin{table}[htb]
\tiny{
\begin{center}
\begin{tabular}{c|c|c|c|c}
\hline
 $e^+e^-\to t q$  &  \multicolumn{2}{c} {$\epsilon_b=60\%$}  & 
 \multicolumn{2}{c} {$\epsilon_b=80\%$} \\
\hline
{\scriptsize ($tq\gamma)$} & $\lambda_{qt}$  &  BR  &  $\lambda_{qt}$  &  BR   \\
\hline
L=0.5 ab$^{-1}$  & 1.8$\times 10^{-02}$   &  1.4$\times 10^{-04}$   &   3.3$\times 10^{-02}$  &  4.7$\times 10^{-04}$   \\
L=10 ab$^{-1}$ & 8.6$\times 10^{-03}$  &  3.2$\times 10^{-05}$   &   1.6$\times 10^{-02}$  &  1.1$\times 10^{-04}$  \\
\hline
\hline
 {\scriptsize ($tqZ)$}$\gamma_\mu$ & $\mathcal{X}_{qt}$  &  BR  &  $\mathcal{X}_{qt}$  &  BR   \\
\hline
L=0.5 ab$^{-1}$  & 2.8$\times 10^{-02}$  &  3.8$\times 10^{-04}$  &  5.1$\times 10^{-02}$  &  1.2$\times 10^{-03}$  \\
L=10 ab$^{-1}$ & 1.3$\times 10^{-02}$  &  8.2$\times 10^{-05}$  &  2.4$\times 10^{-02}$  &  2.8$\times 10^{-04}$ \\
\hline
\hline
{\scriptsize ($tqZ)$}$\sigma_{\mu \nu}$ & $\kappa_{qt}$  &  BR  & $\kappa_{qt}$  &  BR  \\
\hline
L=0.5 ab$^{-1}$   & 2.2$\times 10^{-02}$  &  1.9$\times 10^{-04}$  &   4.1$\times 10^{-02}$  &  6.1$\times 10^{-04}$ \\
L=10 ab$^{-1}$ & 1.1$\times 10^{-02}$  &  4.1$\times 10^{-05}$  &   1.9$\times 10^{-02}$  &  1.4$\times 10^{-04}$ \\
\hline
\end{tabular}
\caption{\small Exclusion limits on FCNC couplings and corresponding BR's at 95\% C.L. from  $e^+e^-\to t q$ at $\sqrt s\simeq 240$~GeV.} 
\end{center}}
\label{tab:one}
\end{table}

\section{EFT approach to top-quark physics at FCC-ee}
\label{sec:Zhang}
{\it Cen Zhang}

At high-energy colliders, precision measurements play an important role in the
search for physics beyond the standard model (SM). In this respect, the top
quark is often considered as a window to new physics, thanks to its large mass.
Results from the LHC measurements already provided valuable information on
possible deviations \cite{Agashe:2014kda}.  Future lepton colliders such as
FCC-ee have been proposed to complement the search program at the LHC, with
much higher sensitivities expected.

A general theoretical framework where the experimental information on possible
deviations from the SM can be consistently and systematically
interpreted is provided by the effective field theory (EFT) approach
\cite{Weinberg:1978kz,Buchmuller:1985jz,Leung:1984ni}.
The EFT Lagrangian corresponds to that of the SM augmented by
higher-dimensional operators that respect the symmetries of the SM. It provides
a powerful approach to identify observables where deviations could be expected
in the top sector, and allows for a global interpretation of all available
measurements. In the following, we briefly discuss several features of the
approach, and the corresponding implications for top physics at future lepton
colliders.

First of all, EFT is a global approach where all possible deviations from the
SM are allowed in the framework, provided that the new physics scale is heavier
than the scale of the experiment. Hence all precision measurements can be
taken into account in this one approach, allowing for a global analysis of the
world's data.  Such an analysis for the top quark has been carried out already
for direct measurements at the Tevatron and the LHC Run-I \cite{Buckley:2015lku,
Buckley:2015nca}.  Other indirect measurements can
be analyzed and combined in the same way, for example those from precision
electroweak data \cite{Zhang:2012cd,deBlas:2015aea} and from flavor data
\cite{Brod:2014hsa}.  As we move on to future colliders, it is important to
continue this program within the same framework, to fully exploit all the
knowledge at hand.

Second, unlike the ``anomalous coupling'' parametrization (see,
e.g.~Ref.~\cite{Beneke:2000hk}) that focuses on
specific vertices, the EFT description is more complete, allowing for
analyzing certain classes of interactions that are not captured in the
``anomalous coupling'' approach, and are often overlooked. For example, the four-fermion interactions
involving two leptons and two quarks, such as
\[O_{et}=(\bar e\gamma^\mu e)(\bar t\gamma_\mu t)\]
are well motivated as they can be easily generated by integrating out a
tree-level heavy mediator in certain new physics scenarios, but are not well
constrained at the LHC.  There are measurements on $t\bar
tl^+l^-$ production which could give some information \cite{Aad:2015eua}, but
the uncertainties
there are quite large, and the process itself is not very sensitive to the
operators.  On the other hand, at future lepton colliders the process
$e^+e^-\to t\bar t$ is much cleaner, and the sensitivity to the operators are
at least two orders of magnitude better.  Another interesting scenario is the
flavor-changing neutral (FCN) interactions coupled to heavy particles.  These
interactions are described by similar operators, such as 
\[O_{eu}^{(a+3)}=(\bar e\gamma^\mu e)(\bar u_a\gamma_\mu t)\]
where $a=1,2$ is the flavor index.  The FCN four-fermion operators have constraints
from $e^+e^-\to tj+\bar tj$ at LEP2 and $t\to jl^+l^-$ at the LHC
\cite{Durieux:2014xla}.  One might think that the LHC gives much tighter bounds on
FCN couplings than LEP2; it turns out that this is true for two-fermion FCN
couplings (such as a $tcZ$ vertex), but is not the case for the four-fermion
ones, mainly because the top-quark three-body decays are suppressed.  This is
reflected in Figure 1, where current bounds on the two types of FCN operators
are compared.  It follows that future lepton collider will provide much more
accurate information on FCN interactions coupled to heavy mediators, and an EFT
approach should be used to capture these interactions and incorporate them
in a global study.

Finally, EFT allows for higher-order
corrections to be consistently added, and therefore predictions can be
systematically improved.  This is important not only for the LHC where QCD
corrections are always large, but also for lepton colliders as the expected
precision level is much higher.  Recently a set of effective operators that
parametrize the chromo-dipole and the electroweak couplings of the top quark
have been implemented in the {\sc MadGraph5\_aMC@NLO} \cite{Alwall:2014hca}
framework at next-to-leading order (NLO) in QCD
\cite{Zhang:2016omx,Bylund:2016phk}.  As a result, starting from an EFT with
these operators, one can make NLO predictions for various top-quark processes,
for cross sections as well as distributions, in a fully automatic way.
Furthermore, NLO results matched to the parton shower simulation are available,
so event generation can be and should be directly employed in future experimental
analyses. These works provide a solid basis for the interpretation of measurements
both at the LHC and at possible future colliders. As an example, in Table 1 we
show the contribution of various operators in $e^+e^-\to t\bar t$ process at a
center-of-mass energy of 500 GeV.

\begin{table}
\begin{center}
\begin{tabular}{ccccccccc}
\hline
   500GeV & SM &  $\mathcal{O}_{tG}$ & $\mathcal{O}^{(3)}_{\phi Q}$ &  $\mathcal{O}^{(1)}_{\phi Q}$ &  $\mathcal{O}_{\phi t}$ &  $\mathcal{O}_{tW}$ &  $\mathcal{O}_{tB}$ \\
\hline
 $\sigma_{i,LO}^{(1)}$  & 566  & 0 & 15.3 & -15.3  & -1.3 & 272  & 191 \\
  $\sigma_{i,NLO}^{(1)}$ &  647 & -6.22 & 18.0 & -18.0 & -1.0 & 307 & 216  \\
 K-factor  &1.14 & N/A  &1.17  & 1.17   & 0.78 & 1.13 & 1.13 \\
 \hline
\end{tabular}
\end{center}
 \caption{\label{tab:sigmattILC} Cross sections (in fb) for $t\bar{t}$
 production at the ILC at $\sqrt{s} =  500$~GeV.  $\sigma_i^{(1)}$ is the
 contribution of operator $\mathcal O_{i}$ at order $\Lambda^{-2}$ assuming
 $\Lambda=1$ TeV.  See Ref.~\cite{Bylund:2016phk} for more details.  }  
\end{table}

\begin{figure}[h!]
\begin{center}
\includegraphics[width=0.70\columnwidth]{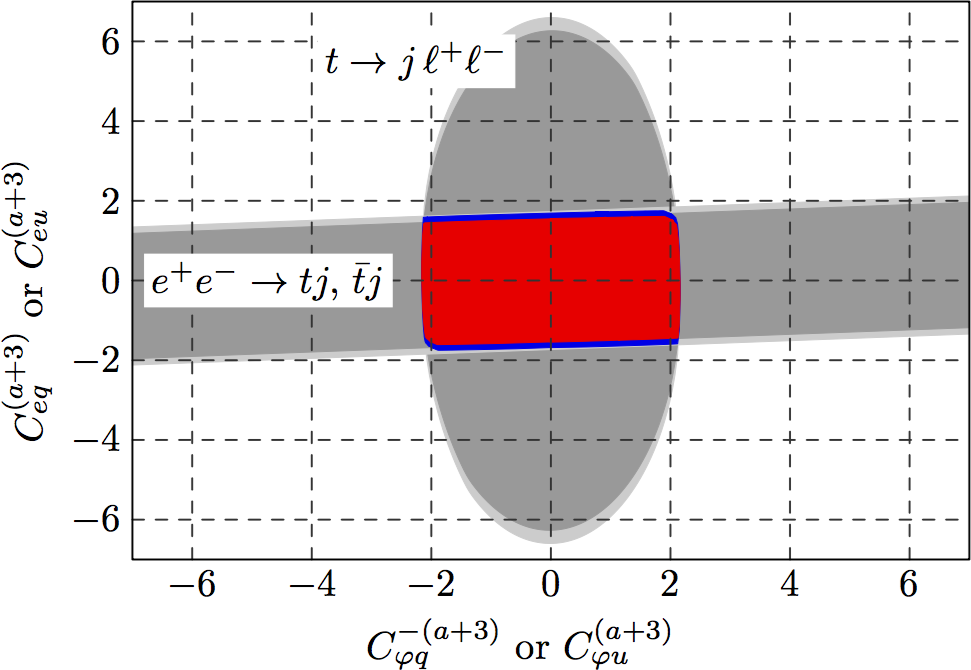}
\caption{Complementarity of the $e^+e^-\to tj+\bar tj$ and $t\to j\,e^+e^-$
limits for constraining two-fermion operators ($O_{\phi q}^{-(a+3)}$
and $O_{\phi u}^{(a+3)}$) and four-fermion ones ($O_{eq}^{(a+3)}$ and
$O_{eu}^{(a+3)}$). The dark gray and red allowed regions apply for $a=1$
while the light gray and blue ones for $a=2$. See
Ref.~\protect\cite{Durieux:2014xla} for more details.%
}
\end{center}
\end{figure}

\section{W mass and width determination using the WW threshold cross section} 
\label{sec:Azzurri}
{\it Paolo Azzurri}

The W mass is a fundamental parameter of the standard model (SM) 
of particle physics, currently measured with a precision of 
15~MeV~\cite{Olive:2016xmw}.
In the context of precision electroweak precision tests
the direct measurement of the W mass is currently limiting
the sensitivity to possible effects of
new physics~\cite{Baak:2014ora}.

A precise direct determination of the W mass can be achieved
by observing the rapid rise of the W-pair production
cross section near its kinematic threshold.
The advantages of this method are that it only involves counting events,
it is clean and uses all decay channels.

In 1996 the LEP2 collider delivered $e^+e^-$ collisions at a single
energy point near 161~GeV, with a total integrated luminosity of
about 10 pb$^{-1}$ at each of the four interaction points.
The data was used to measure the W-pair cross section ($\sigma_{\rm WW}$)
at 161~GeV, and extract the W mass with a precision of 200~MeV
\cite{Barate:1997mn,Abreu:1997sn,Acciarri:1997xc,Ackerstaff:1996nk}.

In the SM the W width is well constrained by the W mass,
and the Fermi constant, with a $~\sim\alpha_S/\pi$ QCD correction
coming from hadronic decays; the W width is currently measured
to a precision of 42~MeV~\cite{Olive:2016xmw}.
The first calculations of the W boson width effects
in $e^+e^-\rightarrow W^+ W^-$ reactions
have been performed in Ref.~\cite{Muta:1986is},
and revealed the substantial effects of the width on the full
cross section lineshape, in particular at energies below the nominal threshold.

In this study the possibility of extracting both the W mass and width
from the determination of $\sigma_{\rm WW}$  at a minimum of two
energy points near the kinematic threshold is explored.

\begin{figure}[h!]
\begin{center}
\includegraphics[width=0.70\columnwidth]{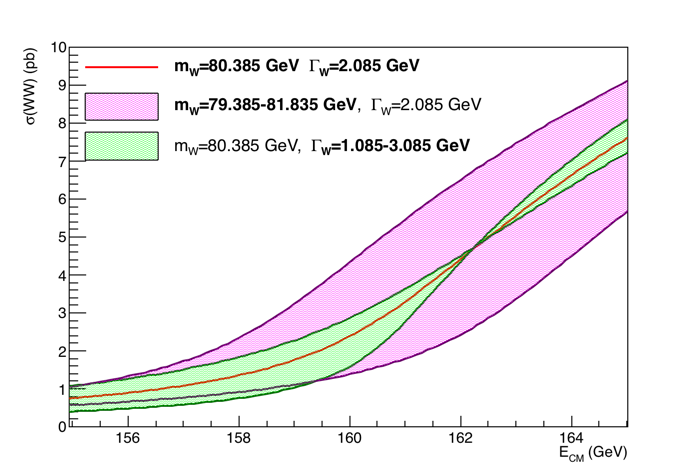}
\caption{W-pair production cross section as a function of
  the $e^+e^-$ collision energy $E_{\rm CM}$.
  The central curve corresponds to the predictions obtained with
  $m_{\rm W}=80.385$~GeV and $\Gamma_{\rm W}=2.085$~GeV.
  Purple and green bands show the cross section curves obtained varying
the W mass and width by $\pm1$~GeV.
\label{fig:sww}
}
\end{center}
\end{figure}

The YFSWW3 version 1.18~\cite{Jadach:2001uu} program has been
used to calculate  $\sigma_{\rm WW}$ as a function of the
energy ($E_{\rm CM}$), W mass ($m_{W^+}$) and width ($\Gamma_{W^+}$).
Figure~\ref{fig:sww} shows the W-pair cross section as a function of the
$e^+e^-$ collision energy with W mass and width values set at the
PDG~\cite{Olive:2016xmw} average measured central values
$m_{\rm W}=80.385$~GeV and $\Gamma_{\rm W}=2.085$~GeV,
and with large 1~GeV variation bands of the mass and width 
central values.

It can be noted that while a variation of the W mass roughly corresponds to a
shift of the cross section lineshape along the energy axis, a variation of the
W width has the effect of changing the slope of cross section lineshape
rise.
It can also be noted that the W width dependence shows a crossing point
at $E_{\rm CM}\simeq 2m_{\rm W} + 1.5 {\rm GeV }\simeq 162.3$~GeV,
where the cross section is insensitive to the W width.

\subsection{W mass measurement at a single energy point}
Detailed studies have illustrated the possibility
of obtaining a precision W mass determination by further
exploiting the threshold W pair production measurement at future 
$e^+e^-$ colliders~\cite{Wilson:2001aw,Wilson:2016hne}.

Taking data at a single energy point the statistical sensitivity
to the W mass with a simple event counting is given by

\begin{equation}
  \Delta m_{\rm W} {\rm (stat)} =
    \left( \frac{d\sigma_{\rm WW}}{dm_{\rm W}}\right)^{-1}
    \frac{\sqrt{\sigma_{\rm WW}}}{\sqrt{\cal L}} \frac{1}{\sqrt{\epsilon p}}
\end{equation}

where ${\cal L}$ is the data integrated luminosity, $\epsilon$
the event selection efficiency and $p$ the selection purity.
The purity can be also expressed as
$$ p = \frac{\epsilon\sigma_{\rm WW}}{\epsilon\sigma_{\rm WW}+\sigma_B}
$$
where $\sigma_B$ is the total selected background
cross section.

A  systematic uncertainty on the background
cross section will propagate to the W mass uncertainty as
\begin{equation}
  \Delta m_{\rm W} {(B)} =
    \left( \frac{d\sigma_{\rm WW}}{dm_{\rm W}}\right)^{-1}
    \frac{\Delta \sigma_B}{\epsilon }.
\end{equation}

Other systematic uncertainties as on the acceptance ($\Delta\epsilon$)
and luminosity ($\Delta{\cal L}$) will propagate as 
\begin{equation}
  \Delta m_{\rm W} {\rm (A)} =
    \left( \frac{d\sigma_{\rm WW}}{dm_{\rm W}}\right)^{-1}
     \left( \frac{\Delta\epsilon}{\epsilon} \oplus
         \frac{\Delta{\cal L}}{\cal L}\right),
\end{equation}
while theoretical uncertainties on the cross section
($\Delta d\sigma_{\rm WW}$) propagate directly as
\begin{equation}
  \Delta m_{\rm W} {\rm (T)} =
  \left( \frac{d\sigma_{\rm WW}}{dm_{\rm W}}\right)^{-1}
  \Delta\sigma_{\rm WW}.
\end{equation}

Finally the uncertainty on the center of mass energy $E_{\rm CM}$
will propagate to the W mass uncertainty as
\begin{equation}
  \Delta m_{\rm W} {\rm (E)} =
  \left( \frac{d\sigma_{\rm WW}}{dm_{\rm W}}\right)^{-1}
  \left( \frac{d\sigma_{\rm WW}}{dE_{\rm CM}}  \right)
  \Delta E_{\rm CM},
\end{equation}
that can be shown to be limited as 
$\Delta m_{\rm W} {\rm (E)} \leq \Delta E_{\rm CM}/2$,
and in fact for $E_{\rm CM}$ near the threshold
it is $\Delta m_{\rm W} {\rm (E)} \simeq \Delta E_{\rm CM}/2$,
so it is the beam energy uncertainty that propagates directly to
the W mass uncertainty.

In the case of ${\cal L}=15$~ab$^{-1}$ accumulated
by the FCCee data taking in one year, and assuming
the LEP event selection quality~\cite{Barate:1997mn}
with $\sigma_B = 300$~fb and $\epsilon = 0.75$,
a statistical precision of $\Delta m_{\rm W}\simeq 0.25$~MeV 
is achievable if the systematic uncertainties will not be limiting the precision,
i.e. if the following conditions are achieved:
\begin{eqnarray}
  \Delta \sigma_B  < 0.5~{\rm fb} \\
  \left( \frac{\Delta\epsilon}{\epsilon} \oplus \frac{\Delta{\cal L}}{\cal L}\right)
  < 10^{-4} \\
  \Delta\sigma_{\rm WW} < 0.6~{\rm fb} \\
  \Delta E_{\rm CM} < 0.25~{\rm MeV}
\end{eqnarray}
corresponding to precision levels of
$10^{-3}$ on the background,
$10^{-4}$ on acceptance and luminosity,
$10^{-4}$ on the theoretical cross section, and 
$0.3\cdot 10^{-5}$ on the beam energy.

\subsection{W mass and width measurements at two energy points}

\begin{figure}[h!]
\begin{center}
\includegraphics[width=0.70\columnwidth]{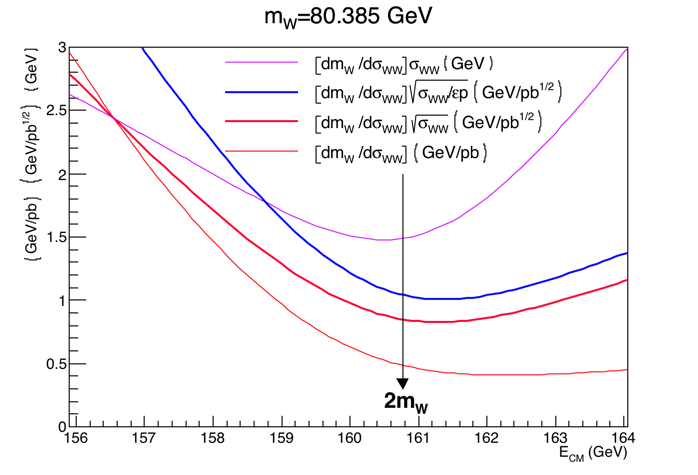}
\caption{\label{fig:dm} W-pair cross section differential functions
  with respect to the W mass.
  The central mass value is set to
  $m_{\rm W}=80.385$~GeV.%
}
\end{center}
\end{figure}

\begin{figure}[h!]
\begin{center}
\includegraphics[width=0.7\columnwidth]{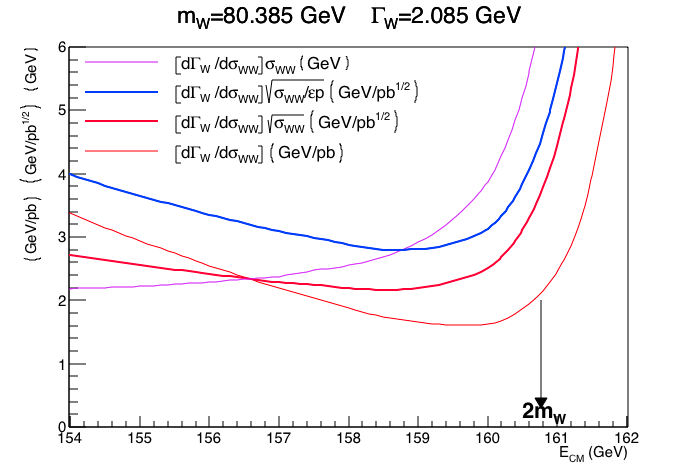}
\caption{{W-pair cross section differential functions
  with respect to the W width.
  Central mass and width values are set to
  $m_{\rm W}=80.385$~GeV and $\Gamma_{\rm W}=2.085$~GeV.\label{fig:dg}}%
}
\end{center}
\end{figure}

Figures~\ref{fig:dm} and \ref{fig:dg} show the differential functions relevant to the
statistical and systematical uncertainties discussed above.
For the statistical terms the efficiency and purities are evaluated
assuming an event selection quality with 
$\sigma_B \simeq  300$~fb and $\epsilon \simeq 0.75$.

The minima of the mass differential curves plotted in Fig.~\ref{fig:dm} 
indicate the optimal points to take data for a W mass measurement,
in particular minimum statistical uncertainty is achieved
with $E_{\rm CM}\simeq 2m_{\rm W} + 0.6$~GeV$\simeq 161.4$~GeV.
The minima of the width differential curves, on Fig.~\ref{fig:dg},
indicate maximum sensitivity to the W width, while all curves diverge
at the W width insensitive point $E_{\rm CM}\simeq 162.3$~GeV, where
$d\sigma_{\rm WW}/d\Gamma_{\rm W}=0$.

If two cross section measurements $\sigma_{1,2}$ are performed at two
energy points $E_{1,2}$, both the W mass and width
can be extracted with a fit to the cross section lineshape.
The uncertainty propagation would then follow
\begin{eqnarray}
  \Delta\sigma_1 = \frac{d\sigma_1}{dm} \Delta m + \frac{d\sigma_1}{d\Gamma} \Delta\Gamma
  = a_1 \Delta m + b_1 \Delta\Gamma \\
  \Delta\sigma_2 = \frac{d\sigma_2}{dm} \Delta m + \frac{d\sigma_2}{d\Gamma} \Delta\Gamma
  = a_2 \Delta m + b_2 \Delta\Gamma.
\end{eqnarray}

The resulting uncertainty on the W mass and width would be
\begin{equation}
  \Delta m = - \frac{b_2\Delta\sigma_1-b_1\Delta\sigma_2}{a_2b_1 - a_1b_2} 
\end{equation}
\begin{equation}
  \Delta \Gamma = \frac{a_2\Delta\sigma_1-a_1\Delta\sigma_2}{a_2b_1 - a_1b_2} 
\end{equation}

If the $\Delta\sigma_{1,2}$ uncertainties on the cross section measurements
are uncorrelated, e.g. only statistical, the linear correlation between
the derived mass and width uncertainties is
\begin{equation}
  r(\Delta m,\Delta \Gamma) = \frac{1}{\Delta m\Delta \Gamma}
  \frac{a_2b_2\Delta\sigma_1^2 + a_1b_1\Delta\sigma_2^2}{(a_2b_1 - a_1b_2)^2}
\end{equation}

\subsection{Optimal data taking configurations}
When conceiving data taking  at two different
energy points near the W-pair threshold
in order to extract both $m_{\rm W}$ and  $\Gamma_{\rm W}$,
it is useful to figure out which energy points values $E_1$ and $E_2$,
would be optimally suited to obtain the best measurements, also
as a function of the data luminosity fraction $f$ delivered at the 
lower energy point.
For this a full 3-dimensional scan of possible $E_1$ and $E_2$ values,
with 100~MeV steps, and of $f$ values, with 0.05 steps, has been performed,
and the data taking configurations that minimize arbitrary combination
of the expected statistical uncertainties on the mass and the width
$f(\Delta m_{\rm W}, \Delta\Gamma_{\rm W})$ are found.

In order to minimize the simple sum of the statistical uncertainties 
$f(\Delta m_{\rm W}, \Delta\Gamma_{\rm W}) = \Delta m_{\rm W} +\Delta\Gamma_{\rm W}$,
the optimal data taking configuration would be with
\begin{eqnarray}
E_1 = 157.5~{\rm GeV},\quad &   E_2 = 162.3~{\rm GeV},\quad &  f=0.45.
\end{eqnarray}
With this configuration, and assuming a total luminosity of
${\cal L}=15$~ab$^{-1}$, the projected statistical uncertainties would be
\begin{eqnarray}
 \Delta m_{\rm W}=0.41~{\rm MeV} \quad &  {\rm and} \quad\Delta \Gamma_{\rm W}=1.10~{\rm MeV}.
\end{eqnarray}
With this same data taking configuration,the statistical uncertainty obtained 
when measuring only the W mass would yield $\Delta m_{\rm W}=0.40$~MeV,
just slightly better with respect to the two-parameter fit.
On the other hand the $\Delta m_{\rm W}=0.40$~MeV precision obtained in this way
must be compared with the $\Delta m_{\rm W}=0.25$~MeV  statistical
precision obtainable when taking all data
at the most optimal single energy point $E_0=161.4$~GeV.

When varying the $f(\Delta m_{\rm W}, \Delta\Gamma_{\rm W})$
target to optimize towards, the obtained optimal energy points
don't change much, with the upper energy always at the
$\Gamma_{\rm W}$-independent $E_2=162.3$~GeV point, and the optimal
lower $E_1$ point at $(1-2)\Gamma_{\rm W}$ units below the nominal
$2m_{\rm W}$ threshold, $E_1=2m_{\rm W} - (1-2) \Gamma_{\rm W}$,
according to if the desired precision is more or less focused
on the W mass or the W width measurement.
In a similar way the optimal data fraction to be taken at the
lower off-shell $E_2$ energy point varies according to the
chosen precision targets, with larger fractions more to the benefit
of the W width precision. When a small fraction of data (e.g. $f=$0.05)
is taken off-shell a statistical precision $\Delta m_{\rm W}=0.28$~MeV
is recovered both in the single- ($ m_{\rm W}$) and the two-parameter
($ m_{\rm W}, \Gamma_{\rm W}$) fits.

\section{Summary}

The goal of the mini-workshop ''Physics behind Precision'' was to provide inspiration for new work on precision physics for FCC-ee in the realm of top quark and electroweak precision physics. The covered topics included Electroweak physics at the Z boson peak, the physics of the $t\bar t$ threshold at $e^+ e^-$ colliders and W boson physics at the $WW$ threshold. The mini-workshop also allowed the possibility of discussing the details of recent event generator developments.  The studies confirm that for most scenarios the FCC-ee will reduce experimental errors by factors of ~10 or more compared to LEP/SLC. This indicates the need for developments in theory to keep the theory uncertainties of similar order.

The mini-workshop succeded in attracting the main proponents from different communities in theory, experiments and tools. The productive discussion during the workshop has resulted in these (online) proceedings. In the future the community aims to further incorporate these and new results on top quark and electroweak precision physics into the FCC-ee Conceptual Design Report under preparation.

\bibliography{converted_to_latex.bib}

\end{document}